\documentclass[aps,pra,twocolumn,superscriptaddress]{revtex4}
\usepackage{bm}
\usepackage{graphicx}
\usepackage{amsmath}
\usepackage{amsthm}
\usepackage{wasysym}
\usepackage[utf8]{inputenc}
\usepackage{amssymb}
\usepackage{epstopdf}
\usepackage{txfonts}
\usepackage{amsfonts}
\begin{document}
\title{Symmetric and asymmetric discrimination of bosonic loss: \\Toy applications to biological samples and photo-degradable materials}
\author{Gaetana Spedalieri}
\affiliation{Computer Science \& York Centre for Quantum
Technologies, University of York, York YO10 5GH, United Kingdom}
\affiliation{Research Laboratory of Electronics, Massachusetts
Institute of Technology, Cambridge, Massachusetts 02139, USA}
\author{Stefano Pirandola}
\affiliation{Computer Science \& York Centre for Quantum Technologies, University of York,
York YO10 5GH, United Kingdom}
\author{Samuel L. Braunstein}
\affiliation{Computer Science \& York Centre for Quantum Technologies, University of York,
York YO10 5GH, United Kingdom}

\begin{abstract}
We consider quantum discrimination of bosonic loss based on both symmetric and
asymmetric hypothesis testing. In both approaches, an entangled resource is
able to outperform any classical strategy based on coherent-state transmitters
in the regime of low photon numbers. In the symmetric case, we then consider
the low-energy detection of bacterial growth in culture media. Assuming an
exponential growth law for the bacterial concentration and the Beer-Lambert
law for the optical transmissivity of the sample, we find that the use of
entanglement allows one to achieve a much faster detection of growth with
respect to the use of coherent states. This performance is also studied by
assuming an exponential photo-degradable model, where the concentration is
reduced by increasing the number of photons irradiated over the sample. This
investigation is then extended to the readout of classical information from
suitably-designed photo-degradable optical memories.

\end{abstract}
\maketitle

\section{Introduction}

Today we use classical or semi-classical sources of radiation for a plethora
of practical applications. A natural question to ask is how far we can improve
these applications by resorting to the laws of quantum mechanics and the new
framework of quantum
information~\cite{Watrous,Hayashi,NiCh,SamRMPm,RMP,AdessoR,sera,hybrid}. In
fact, nonclassical states represent a completely new resource for future
quantum technology, once their generation and manipulation becomes routinely
possible outside quantum labs. Examples of nonclassical states are number
states, squeezed states and entangled states. In particular, quantum
entanglement is one of the key properties exploited in many quantum
information tasks, e.g., in protocols of quantum
teleportation~\cite{tele,teleCV,telereview}, quantum
metrology~\cite{SamMETRO,Caves,Paris,Giova,ReviewMETRO,ReviewNEW}, and quantum
cryptography~\cite{Ekert}. In the bosonic case, entanglement is usually
present in the form of Einstein-Podolsky-Rosen (EPR)
correlations~\cite{EPRpaper}, where the quadrature operators of two bosonic
modes are so correlated to beat any possible classical description.

Entanglement and, more generally, quantum correlations have already led to
non-trivial improvements in information tasks that are based on quantum
hypothesis testing~\cite{QHT,QHT1,QHT2,UNA1,UNA2,UNA3,UNA4,UNA5,Invernizzi},
such as the detection of targets or quantum
illumination~\cite{Qill0,Qill1,Qill1b,Qill2,Qill3,Qillexp1,Qillexp2,Qillexp3,PBTpaper}%
, and the readout of digital optical memories or quantum
reading~\cite{Qread,Qread2,Qread3,Qread4,Qread4b,Qread5,Qread6,Qread7,Qread7b,Qread7c,Qread8,Qread9,Qread9b,Qread10}%
. These enhancements are particularly evident when the number of photons
involved in the process is relatively low. A very important scenario where low
energy matters is in the context of fragile material in biological samples. In
this case, faint quantum light may effectively probe the material without
destroying it.

In this paper, we consider this context. We study the detection of bosonic
loss in fragile photo-degradable materials. First, we review the theory of
quantum discrimination of bosonic loss by extending previous results from
symmetric to asymmetric hypothesis testing. Then, we show how quantum
entanglement can effectively be used to discriminate the presence or absence
of bacterial growth in a biological sample. The advantage is quantified in
terms of speed of detection with respect to the use of a coherent-state
transmitter under the same (relatively-low) number of mean photons irradiated
over the sample. We show energy regimes where this advantage becomes
non-trivial, and how it can also be used to build fragile memories where
information remains hidden to standard optical readers (extending previous
ideas from Ref.~\cite{Qread9b}).

The manuscript is structured as follows. In Sec.~\ref{sec1} we provide a brief
review of quantum hypothesis testing considering its symmetric and asymmetric
formulation. In Sec.~\ref{sec2}, we consider the quantum discrimination of
bosonic loss and its connection with the detection of non-zero concentration
in samples. In Sec.~\ref{sec3},\ we discuss the classical benchmark associated
with coherent-state transmitters. While this is known for symmetric testing,
here we also provide its expression for the asymmetric case. In
Sec.~\ref{sec4}, we then consider the entanglement-based quantum transmitter,
providing its performance for both symmetric and asymmetric testing. In
Sec.~\ref{sec5}, we explicitly compare classical and quantum transmitters
under the two different types of testing, and consider single- and multi-copy
(asymptotic) discrimination. In Sec.~\ref{sec6}\ we apply the tools to
bacterial growth in a sample assuming an exponential growth model and a toy
model of photo-degradability. In Sec.~\ref{sec7}, we extend the treatment to
digital memories. Finally, Sec.~\ref{sec8} is for conclusions.

\section{Quantum hypothesis testing\label{sec1}}

\subsection{Bayesian cost\label{APP_Bayes}}

We start by providing a brief survey of the concept of Bayesian cost and its
connection with symmetric and asymmetric hypothesis testing. For simplicity,
we consider the specific case of a binary hypothesis test, but the notion can
easily be generalized and formulated for $N\geq2$ different hypotheses.

Let us consider a binary test with hypotheses $H_{0}$ (null) and $H_{1}$
(alternate), occurring with some a priori probabilities $p_{0}$ and $p_{1}$,
respectively. In the quantum setting, these hypotheses are associated with two
possible states, $\rho_{0}$ and $\rho_{1}$, taken by some quantum system,
i.e., we may write
\begin{align}
H_{0}  &  :\rho=\rho_{0}~,\\
H_{1}  &  :\rho=\rho_{1}~.
\end{align}

Associated with the test, there is the conditional probability $p(k|i)$ of
choosing the hypothesis $H_{k}$ when the actual hypothesis is $H_{i}$, with
$i,k\in\{0,1\}$. In a quantum test, this probability is determined by the
quantum measurement which is performed on the system. In particular, it is
always sufficient to consider a dichotomic quantum measurement, described by a
positive operator-valued measure (POVM) with two operators $M_{0}\geq0$ and
$M_{1}=I-M_{0}$. Thus, we may write the Born rule%
\begin{equation}
p(k|i)=\mathrm{Tr}\left(  M_{k}\rho_{i}\right)  ~.
\end{equation}

For $k\neq i$, we have error probabilities, known as false-negative
probability $p(0|1)$, and false-positive probability $p(1|0)$. In general, we
can also introduce a cost matrix $\mathbf{C}$, whose generic element $C_{ki}$
represents the `cost' associated with conditional probability $p(k|i) $. Thus,
the goal of the binary test is to minimize the following Bayes' cost function%
\begin{equation}
\mathcal{C}_{\text{B}}:=\sum_{i,k}C_{ki}~p_{i}~p(k|i)~.
\end{equation}
In particular, we may choose
\begin{equation}
\mathbf{C}=\left(
\begin{array}
[c]{cc}%
0 & C_{01}\\
C_{10} & 0
\end{array}
\right)  ~,
\end{equation}
so that
\begin{equation}
\mathcal{C}_{\text{B}}=C_{10}p_{0}p(1|0)+C_{01}p_{1}p(0|1).
\end{equation}

Depending on the choice of $C_{01}$ and $C_{10}$, we may have a symmetric or
asymmetric hypothesis test. When the costs are the same ($C_{01}=C_{10}=1$),
we retrieve symmetric hypothesis testing. Here the cost function simply
becomes the mean error probability of the test
\begin{equation}
\mathcal{C}_{\text{B}}=\bar{p}:=p_{0}p(1|0)+p_{1}p(0|1).\label{avERRORprob}%
\end{equation}
By contrast, in case of completely unbalanced costs, such as $C_{01}=1$ or
$C_{10}=0$, we have asymmetric hypothesis testing, with the cost function
collapsing into the false-negative error probability, i.e., $\mathcal{C}%
_{\text{B}}=p(0|1)$.

\subsection{Symmetric Testing}

Here we describe the main tools used in the symmetric testing of two quantum
hypotheses, considering (for simplicity) the case of a uniform a priori
probability distribution $p_{0}=p_{1}=1/2$. The mean error probability
associated with the test is minimized by the Helstrom POVM~\cite{QHT}~ and is
given by the Helstrom bound%
\begin{equation}
\bar{p}=\frac{1-D(\rho_{0},\rho_{1})}{2},
\end{equation}
where $D(\rho_{0},\rho_{1})$ is the trace distance between the two quantum
states~\cite{Watrous,Hayashi}.

In general, the two states can be a tensor product of $M$ single-copy states,
i.e.,%
\begin{equation}
\rho_{i}=\sigma_{i}^{\otimes M}:=\underset{M}{\underbrace{\sigma_{i}%
\otimes\cdots\otimes\sigma_{i}}}~.
\end{equation}
In this multi-copy discrimination setting, a very useful tool is the quantum
Chernoff bound (QCB)~\cite{QCbound,QCbound2,MinkoPRA}. This is an upper-bound
$\bar{p}_{\text{QCB}}\geq\bar{p}$, which is defined as%
\begin{equation}
\bar{p}_{\text{QCB}}:=\frac{C^{M}}{2},~~~~C:=\inf_{s\in(0,1)}C_{s}%
,\label{QCBdef}%
\end{equation}
where $C_{s}:=\mathrm{Tr}(\sigma_{0}^{s}\sigma_{1}^{1-s})$ is the $s$-overlap
between the two single-copy states $\sigma_{0}$ and $\sigma_{1}$. The QCB is
asymptotically tight, meaning that we have $\bar{p}_{\text{QCB}}\simeq\bar{p}$
for $M\rightarrow+\infty$~\cite{QCbound,QCbound2}.

Other computable bounds can be considered satisfying
\begin{equation}
\bar{p}_{\text{F}}\leq\bar{p}\leq\bar{p}_{\text{QCB}}\leq\bar{p}_{\text{QBB}%
}~,
\end{equation}
where
\begin{equation}
\bar{p}_{\text{QBB}}:=\frac{1}{2}C_{1/2}^{M} \label{QBBdef}%
\end{equation}
is the quantum Battacharyya bound (QBB)~\cite{MinkoPRA} and~we
have~\cite{Fuchs}%
\begin{equation}
\bar{p}_{\text{F}}:=\frac{1-\sqrt{1-F^{M}}}{2}~, \label{QFBdef}%
\end{equation}
with
$F:=[\mathrm{Tr}\sqrt{\sqrt{\sigma_{0}}\sigma_{1}\sqrt{\sigma_{0}}}]^{2}$
being the single-copy fidelity~\cite{fid1,fid2} between
$\sigma_{0}$ and $\sigma _{1}$.

In the specific case where $\sigma_{0}$ and $\sigma_{1}$ are Gaussian states,
one can easily compute their fidelity $F$ and their overlap $C_{s}$, starting
from their statistical moments~\cite{MinkoPRA,Spedalieri13}. One can therefore
derive all the above bounds $\bar{p}_{\text{F}}$, $\bar{p}_{\text{QCB}}$, and
$\bar{p}_{\text{QBB}}$ (see Appendix~\ref{Gauss_APP}).

\subsection{Asymmetric Testing}

In the regime of many copies $M$, we can write the following asymptotic
behaviors for the error probabilities associated with the test%
\begin{align}
p(1|0)  &  \simeq\frac{1}{2}\exp(-\alpha_{R}M)~,\\
p(0|1)  &  \simeq\frac{1}{2}\exp(-\beta_{R}M)~,
\end{align}
where $\alpha_{R}$ is the error-rate exponent of the false-positive
probability, and $\beta_{R}$ is the error-rate exponent of the false-negative probability.

In asymmetric quantum discrimination, we aim to maximize the error-rate
exponent of the false negatives $\beta_{R}$ while constraining the error-rate
exponent of the false positives $\alpha_{R}\geq r$ with a positive parameter
$r$. By optimizing over all the possible quantum measurements $\mathcal{M}$,
the maximum error-rate $\underset{\mathcal{M}}{\max}\beta_{R}$ is given by the
quantum Hoeffding bound (QHB)~\cite{QHB}, which is defined as%
\begin{equation}
H(r)=\sup_{0\leq s<1}P(r,s),~~~~P(r,s):=\frac{-rs-\ln C_{s}}{1-s}%
.\label{QHBdef}%
\end{equation}
In case of Gaussian states $\sigma_{0}$ and $\sigma_{1}$, the QHB can be
computed from their statistical moments~\cite{GaeQHB} (see
Appendix~\ref{Gauss_APP}).

\subsection{Discriminating a pure state from a mixed state\label{SEConePoneM}}

In the specific case where one of the two states is pure, say $\sigma
_{0}=\left\vert \varphi\right\rangle \left\langle \varphi\right\vert $, we
have some simplifications in the above tools. First of all, the QCB can be
directly computed from the quantum fidelity as
\begin{equation}
C=F=\left\langle \varphi\right\vert \sigma_{1}\left\vert \varphi\right\rangle
,
\end{equation}
so that we can write
\begin{equation}
\bar{p}_{\text{QCB}}=\frac{F^{M}}{2}~.\label{QCBsimpleFID}%
\end{equation}

Then, the QHB satisfies the inequality%
\begin{equation}
H(r)\geq-\ln F~.\label{hb1}%
\end{equation}
As a result, we may write
\begin{equation}
p(0|1)\simeq\frac{1}{2}\exp(-HM)\leq\bar{p}_{\text{QCB}}~.
\end{equation}
This is intuitively expected because it is surely easier to minimize the
false-negative error probability $p(0|1)$, constraining the false positive
error probability $p(1|0)$, rather than minimizing their average\ $\bar{p}$.
However, if the control on the false-positives is good, here meaning that
$r\geq-\ln F$, then the QHB and the QCB coincides. In fact, we have%
\begin{equation}
H(r)=-\ln F~~\text{for~~}r\geq-\ln F,\label{hb2}%
\end{equation}
which clearly implies $p(0|1)\simeq\bar{p}_{\text{QCB}}$.

The proof of Eqs.~(\ref{hb1}) and (\ref{hb2}) is easy. First, from the
definition of the QHB, we note that we can generally write the inequality%
\begin{align}
H(r)  &  \geq\lim_{s\rightarrow0}P(r,s)=\lim_{s\rightarrow0}\left(  -\ln
C_{s}\right) \\
&  =-\ln\left(  \lim_{s\rightarrow0}C_{s}\right)  ~.
\end{align}
Now, if $\sigma_{0}=\left\vert \varphi\right\rangle \left\langle
\varphi\right\vert $, then we have~\cite{Spedalieri13}
\begin{equation}
\lim_{s\rightarrow0}C_{s}=F\left(  \left\vert \varphi\right\rangle ,\sigma
_{1}\right)  ~,
\end{equation}
which provides Eq.~(\ref{hb1}). Furthermore, in this case of one pure state,
Ref.~\cite{GaeQHB} proved the following upper-bound%
\begin{equation}
H(r)\leq H_{F}(r):=\left\{
\begin{array}
[c]{l}%
-\ln F~~\text{for~~}r\geq-\ln F~,\\
\\
+\infty~~\text{for~~}r<-\ln F~.
\end{array}
\right.  \label{hb3}%
\end{equation}
By combining Eq.~(\ref{hb1}) and~(\ref{hb3}), we then derive the result of
Eq.~(\ref{hb2}).

\subsection{Discriminating two pure states\label{Sec_twopure}}

In the stronger case where both the states are pure, i.e., $\sigma
_{0}=\left\vert \varphi_{0}\right\rangle \left\langle \varphi_{0}\right\vert $
and $\sigma_{1}=\left\vert \varphi_{1}\right\rangle \left\langle \varphi
_{1}\right\vert $, we have further simplifications. For the multi-copy
Helstrom bound, we can write%
\begin{equation}
\bar{p}=\frac{1-D\left(  \left\vert \varphi_{0}\right\rangle ^{\otimes
M},\left\vert \varphi_{1}\right\rangle ^{\otimes M}\right)  }{2},
\label{pmedio}%
\end{equation}
where the trace distance $D$ can be here computed from the fidelity as%
\begin{align}
D  &  =\sqrt{1-F\left(  \left\vert \varphi_{0}\right\rangle ^{\otimes
M},\left\vert \varphi_{1}\right\rangle ^{\otimes M}\right)  }\label{traceD}\\
&  =\sqrt{1-F\left(  \left\vert \varphi_{0}\right\rangle ,\left\vert
\varphi_{1}\right\rangle \right)  ^{M}}, \label{fid}%
\end{align}
where
\begin{equation}
F(\left\vert \varphi_{0}\right\rangle ,\left\vert \varphi_{1}\right\rangle
)=\left\vert \left\langle \varphi_{0}\right.  \left\vert \varphi
_{1}\right\rangle \right\vert ^{2}~.
\end{equation}

Then, for the QHB we can write$~$\cite{GaeQHB}
\begin{equation}
H(r)=H_{F}(r)~. \label{Hf2}%
\end{equation}
In particular, for bad control on the false positives ($r<-\ln F$) we have
$H(r)=+\infty$, which means that the asymptotic decay of the false-negative
error probability $p(0|1)$ is super-exponential~\cite{GaeQHB}.

\section{Quantum discrimination of Loss\label{sec2}}

The above tools of quantum hypothesis testing can also be employed to solve
problems of quantum channel discrimination. Here, an unknown quantum channel,
$\mathcal{E}_{0}$ or $\mathcal{E}_{1}$, is prepared inside an input-output
black-box and passed to a reader~\cite[Sec.~V.H]{RMP}, whose aim is to
distinguish the two channels by probing the input port and measuring the
output. In the bosonic setting, this problem can be further constrained by
considering Gaussian channels~\cite{RMP}, in particular, lossy channels
$\mathcal{E}_{\tau}$ which are characterized by a single transmissivity
parameter $\tau\in\lbrack0,1]$. These channels can be dilated into
beam-splitters subject to vacuum noise (see Appendix~\ref{APP_lossy} for more
details on lossy channels).

In this paper we are interested in sensing the presence of loss, which
corresponds to the discrimination between a lossless channel (i.e., the
identity channel $\mathcal{I}$) from a lossy channel with some transmissivity
$\tau<1$. In other words, we consider transmissivity $\tau_{0}=1$ as our null
hypothesis $H_{0}$, and transmissivity $\tau_{1}:=\tau<1$ as our alternative
hypothesis $H_{1}$.

This problem is relevant in biological problems, such as the detection of
small concentrations of cells or bacteria. The connection is established by
the Beer-Lambert law~\cite{Ingle88}. Accordingly to this law, the
concentration $c$ of species within a sample can be connected with its
absorbance or transmissivity $\tau$ by the formula%
\begin{equation}
\tau=10^{-\varepsilon lc},\label{LB1}%
\end{equation}
where $\varepsilon$\ is an absorption coefficient (for that species), and $l$
is the path length. From an optical point of view, the sample is therefore
equivalent to a lossy channel with concentration-dependent transmissivity
$\tau=\tau(c)$. Our problem can be mapped into the discrimination between
non-growth $(c=0)$ and growth $(c>0)$ within a biological sample.

An important issue is related with the amount of energy employed to probe the
black-box or, equivalently, the sample in the biological setting just
described. First of all, a problem of Gaussian channel discrimination makes
sense only if we assume an energetic constraint for the optical mode probing
the box, otherwise any two different channels can always be perfectly
distinguished (using infinite energy).

Second, we assume that this constraint imposes an effective regime of few
photons, so that the box is read in a noninvasive way, which fully preserves
its content. This is particularly important from a biological point of view,
since bacteria may be photo-degradable and DNA/RNA extracts in samples can
easily be degraded by strong light (especially, in the UV regime).

Thus, in our quantum sensing of loss, we assume suitable energetic constraints
at the input, which may be of two kinds:

\begin{description}
\item[(1)] Local energetic constraint, where we fix the mean number of photons
employed in each single readout of the box; in particular, we are interested
in the use of a single readout and in the limit of many readouts (e.g., using
a broadband signal).

\item[(2)] Global energetic constraint, where we fix the mean number of
photons which are used in totality, i.e., over all the readouts of the box.
\end{description}

\noindent Imposing one of these constraints and a suitable regime of few
photons, our work aims to prove the superiority of quantum-correlated sources
with respect to classical sources for noninvasive sensing of loss.

\begin{figure}[ptbh]
\vspace{-0.2cm}
\par
\begin{center}
\includegraphics[width=0.5\textwidth] {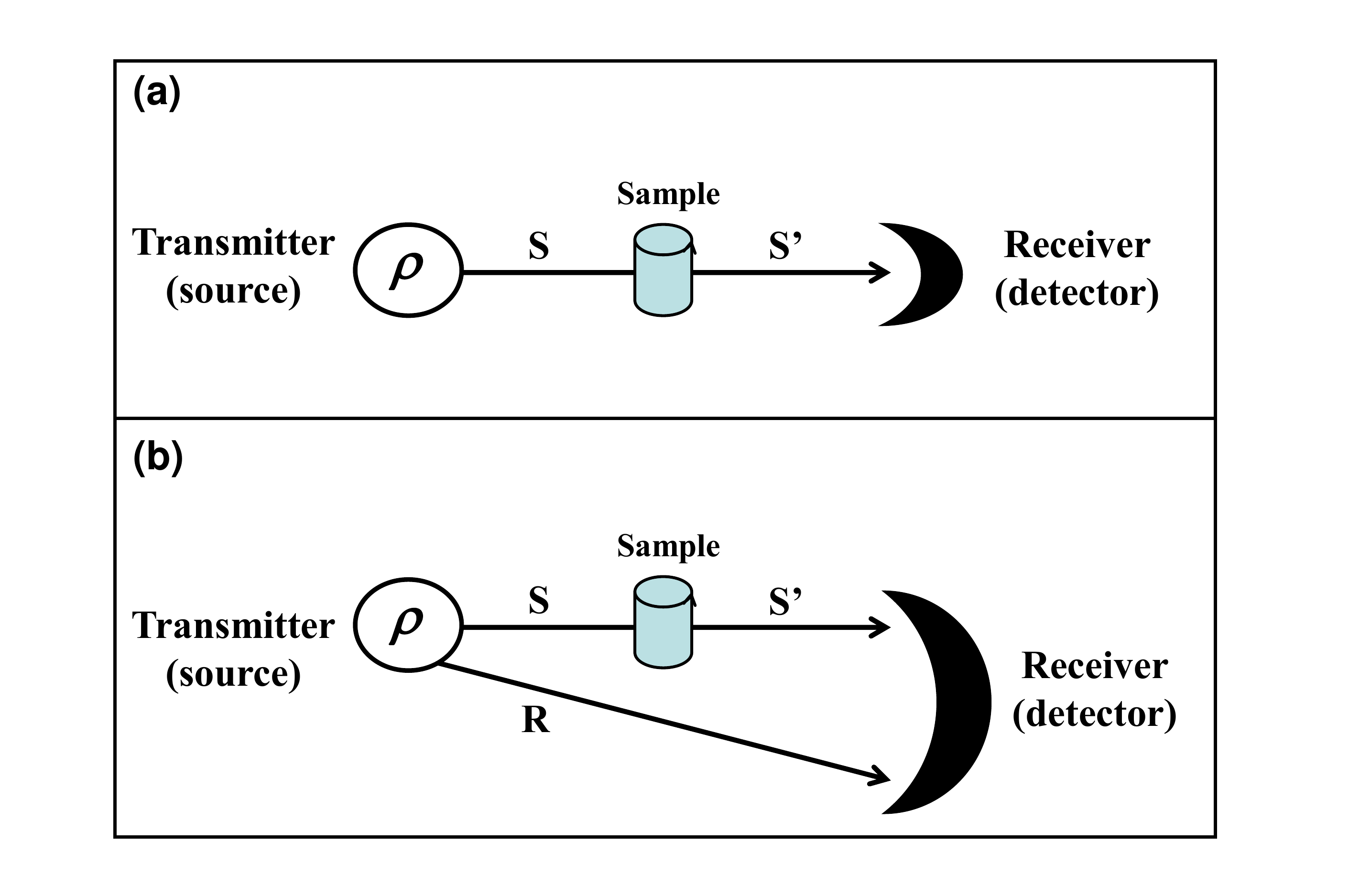}
\end{center}
\par
\vspace{-0.4cm}\caption{Configurations for sensing the presence of loss in a
sample via a transmitter (source) and a receiver (detector).
\textbf{Panel~(a)}. In the classical setup, a signal mode $S$ is prepared in a
coherent state and irradiated through the sample, with the output mode
$S^{\prime}$ subject to optimal detection.\textbf{\ Panel~(b)}. In the quantum
setup, the transmitter is composed of two quantum-correlated modes, $S$ and
$R$. Only $S$ is irradiated through the sample. The output $S^{\prime}$ is
combined with $R$ in a joint optimal quantum measurement.}%
\label{Figure}%
\end{figure}

As shown by the setups of Fig.~\ref{Figure}, we first consider a classical
strategy where a single-mode $S$, prepared in a classical state (in
particular, a coherent state), is irradiated through the sample and detected
at the output by an optimal dichotomic POVM. Then, we compare this approach
with the quantum strategy where two modes, signal $S$ and reference $R$, are
prepared in a quantum correlated state, in particular, a two-mode squeezed
vacuum (TMSV) state, which is a finite-energy version of an
Einstein-Podolsky-Rosen (EPR) state~\cite{RMP}. Only the signal mode is
irradiated through the sample, while the reference mode is directly sent to
the measurement setup where it is subject to an optimal dichotomic POVM
jointly with the output mode $S^{\prime}$ from the sample.

The readout performance of these two setups are compared by constraining the
energy of the signal mode $S$, by fixing the mean number of photon $\bar{n}$
per mode (local constraint), or the total mean number of photons $\bar
{N}=M\bar{n}$ in $M$ probings of the sample. The performance is evaluated in
terms of minimum error probability considering both symmetric and asymmetric
hypothesis testing. 

\section{Classical Benchmark\label{sec3}}

In the classical setup of Fig.~\ref{Figure}(a), the input signal mode $S$ is
prepared in a coherent state $\left\vert \alpha\right\rangle $\ and
transmitted through the sample. At the output, the receiver may be a photon
counting detector, which is then followed by post-processing of the outcomes.
The performance of this receiver can always be bounded by considering an
optimal dichotomic POVM (e.g., the Helstrom POVM~\cite{QHT} in symmetric
testing and other suitable dichotomic POVMs in the asymmetric case~\cite{QHB}).

Let us solve this problem in the general multi-copy scenario, where the sample
is probed $M$ times, so that the input state is given by the tensor product%
\begin{equation}
\left\vert \alpha\right\rangle ^{\otimes M}=\underset{M}{\underbrace
{\left\vert \alpha\right\rangle \otimes\cdots\otimes\left\vert \alpha
\right\rangle }}~.
\end{equation}
It is clear that the output states will be either $\left\vert \alpha
\right\rangle ^{\otimes M}$ (under hypothesis $H_{0}$)\ or $\left\vert
\sqrt{\tau}\alpha\right\rangle ^{\otimes M}$ (under hypothesis $H_{1}$).

Since the two possible outputs are pure states, we can easily compute the
Helstrom bound (for symmetric testing) and the QHB (for asymmetric testing)
according to Sec.~\ref{Sec_twopure}. In fact, we just need to compute the
fidelity between the single-copy output states%
\begin{align}
F\left(  \left\vert \alpha\right\rangle ,\left\vert \sqrt{\tau}\alpha
\right\rangle \right)   &  =\exp\left(  -\left\vert \alpha-\sqrt{\tau}%
\alpha\right\vert ^{2}\right)  \label{fidcoh}\\
&  =\exp[-\bar{n}(1-\sqrt{\tau})^{2}],
\end{align}
where $\bar{n}=\left\vert \alpha\right\vert ^{2}$ is the mean number of
photons of the single-copy coherent state at the input.

Then, using Eqs.~(\ref{pmedio})-(\ref{fid}), one derives the following
Helstrom bound for the coherent-state transmitter~\cite{Qread}
\begin{equation}
\bar{p}_{\text{coh}}=\frac{1-\sqrt{1-e^{-\bar{N}(1-\sqrt{\tau})^{2}}}}%
{2}.\label{CLASStransmitter}%
\end{equation}
We can see that the minimum error probability depends on the total mean number
of photons $\bar{N}=M\bar{n}$. This means that it makes no difference to use:
(i) either $M$ identical faint coherent states each with $\bar{n}$ mean
photons, (ii) or a single energetic coherent state with $\bar{N}$ mean
photons. Also note that, for $\bar{N}(1-\sqrt{\tau})^{2}\ll1$, we have
$\bar{p}_{\text{coh}}\simeq1/2$, i.e., random guessing. Discrimination becomes
therefore challenging at low photon numbers.

Note that it is also easy to compute the QCB. Since the two states are pure,
we can write it in terms of the quantum fidelity as specified by
Eq.~(\ref{QCBsimpleFID}), which here becomes%
\begin{equation}
\bar{p}_{\text{coh}}^{\text{QCB}}=\frac{1}{2}\exp[-\bar{N}(1-\sqrt{\tau}%
)^{2}].\label{QCB1}%
\end{equation}
One can easily check that this is an upper-bound to the minimum error
probability of Eq.~(\ref{CLASStransmitter}), becoming tight in the limit of a
large number of copies $M\gg1$.

In the case of asymmetric quantum discrimination, we aim to minimize the
probability of false negatives $p(0|1)$. In a biological sample, this means
one should minimize the probability of concluding that there is no growth of
bacteria when there actually is. More precisely, we aim to derive the QHB
which maximizes the error-rate exponent $\beta_{R}$ of $p(0|1)$ in the regime
of many copies $M$, while constraining the error-rate exponent for the false
positives $\alpha_{R}\geq r$. By using Eq.~(\ref{Hf2}), we derive%
\begin{equation}
H_{\text{coh}}(r)=\left\{
\begin{array}
[c]{l}%
-\ln F=\bar{n}(1-\sqrt{\tau})^{2},~~\text{for~~}r\geq\bar{n}(1-\sqrt{\tau
})^{2}\\
\\
+\infty,~~~\text{for~~}r<\bar{n}(1-\sqrt{\tau})^{2}~~~~~~~~~~~~~~~~~
\end{array}
\right.  \label{QHBcoherent}%
\end{equation}

Here we note that for bad control of the false positives $r<\bar{n}%
(1-\sqrt{\tau})^{2}$, the QHB has a super-exponential decay in $M$. In
contrast, for good control of the false positives $r\geq\bar{n}(1-\sqrt{\tau
})^{2}$, the QHB has the following asymptotic exponential decay%
\begin{align}
p_{\text{coh}}(0|1) &  \simeq\frac{1}{2}\exp[-M~H(r)]\\
&  \simeq\frac{1}{2}\exp\left[  -\bar{N}(1-\sqrt{\tau})^{2}\right]
,\label{QHB2}%
\end{align}
which is the same as the QCB in Eq.~(\ref{QCB1}). This an intuitive result
because in the case of good control, besides Eq.~(\ref{QHB2}), we also have%
\begin{align}
p_{\text{coh}}(1|0) &  \simeq\frac{1}{2}\exp[-M~r]\\
&  \leq\frac{1}{2}\exp\left[  -\bar{N}(1-\sqrt{\tau})^{2}\right]
.\label{QHB3}%
\end{align}
Thus, by replacing Eqs.~(\ref{QHB2}) and~(\ref{QHB3}) in the average error
probability of Eq.~(\ref{avERRORprob}), we retrieve
\begin{equation}
\bar{p}_{\text{coh}}\lesssim\frac{1}{2}\exp\left[  -\bar{N}(1-\sqrt{\tau}%
)^{2}\right]  .
\end{equation}

\section{Quantum Transmitter\label{sec4}}

In the quantum setup of Fig.~\ref{Figure}(b), we consider a transmitter
composed of two quantum-correlated modes, the signal $S$ and the reference $R
$. The signal mode, with $\bar{n}$ mean photons, is irradiated through the
sample and its output $S^{\prime}$ is combined with the reference mode in an
optimal quantum measurement. For a fixed state $\rho_{SR}$ of the input modes
$S$ and $R$, we get two possible states
\begin{align}
\sigma_{0}  &  =(\mathcal{I}\otimes\mathcal{I})(\rho_{SR})=\rho_{SR},\\
\sigma_{1}  &  =(\mathcal{E}_{\tau}\otimes\mathcal{I})(\rho_{SR}),
\end{align}
for the output modes $S^{\prime}$ and $R$ at the receiver. In general, for
multi-copy discrimination, the input tensor product $\rho_{SR}^{\otimes M}$ is
transformed into either $\sigma_{0}^{\otimes M}$ or $\sigma_{1}^{\otimes M}$.

As an example of a single-copy state $\rho_{SR}$ let us consider a TMSV state
(or realistic EPR\ source). This is a zero-mean pure Gaussian state
$\left\vert \mu\right\rangle _{SR}$ with covariance matrix (CM)~\cite{RMP}%
\begin{equation}
\mathbf{V}(\mu)=\left(
\begin{array}
[c]{cc}%
\mu\mathbf{I} & \sqrt{\mu^{2}-1}\mathbf{Z}\\
\sqrt{\mu^{2}-1}\mathbf{Z} & \mu\mathbf{I}%
\end{array}
\right)  ,~%
\begin{array}
[c]{c}%
\mathbf{Z}:=\mathrm{diag}(1,-1),\\
\mathbf{I}:=\mathrm{diag}(1,1),~~
\end{array}
\end{equation}
where $\mu\geq1$ quantifies both the mean number of thermal photons in each
mode, given by $\bar{n}=(\mu-1)/2$, and the amount of signal-reference
entanglement~\cite{RMP}.

Using such a state at the input, we get two possible zero-mean Gaussian states
at the output: One is just the input TMSV state $\sigma_{0}=\left\vert
\mu\right\rangle _{SR}\left\langle \mu\right\vert $, while the other state
$\sigma_{1}$ is mixed and has CM (see Appendix~\ref{APP_CM})%
\begin{equation}
\mathbf{V}_{1}(\mu,\tau)=\left(
\begin{array}
[c]{cc}%
(\tau\mu+1-\tau)\mathbf{I} & \sqrt{\tau(\mu^{2}-1)}\mathbf{Z}\\
\sqrt{\tau(\mu^{2}-1)}\mathbf{Z} & \mu\mathbf{I}%
\end{array}
\right)  .\label{CM1out}%
\end{equation}
Since one of the output states is pure, we can exploit the tools of
Sec.~\ref{SEConePoneM}.

For symmetric testing we compute the QCB, which can be directly derived from
the quantum fidelity, as specified in Eq.~(\ref{QCBsimpleFID}). The multi-copy
minimum error probability $\bar{p}_{\text{quant}}$ is upper-bounded by
\begin{equation}
\bar{p}_{\text{quant}}^{\text{QCB}}=\frac{F^{M}}{2},
\end{equation}
where the fidelity $F=\left\langle \mu\right\vert \sigma_{1}\left\vert
\mu\right\rangle $ is determined by the CMs of the two Gaussian
states~\cite{Spedalieri13}. This is equal to%
\begin{equation}
F=\frac{4}{\sqrt{\det[\mathbf{V}(\mu)+\mathbf{V}_{1}(\mu,\tau)]}}=\left[
1+\bar{n}\left(  1-\sqrt{\tau}\right)  \right]  ^{-2}.\label{Fgauss2}%
\end{equation}
Thus, we have~\cite{Qread}%
\begin{equation}
\bar{p}_{\text{quant}}\leq\bar{p}_{\text{quant}}^{\text{QCB}}=\frac{1}%
{2}\left[  1+\bar{n}\left(  1-\sqrt{\tau}\right)  \right]  ^{-2M}%
.\label{EPRtransmitter}%
\end{equation}

For asymmetric testing we compute the QHB. From Eq.~(\ref{hb1}), we may write%
\begin{equation}
H_{\text{quant}}(r)\geq-\ln F=2\ln\left[  1+\bar{n}\left(  1-\sqrt{\tau
}\right)  \right]  .\label{QHBquantumEPR}%
\end{equation}
More precisely, according to Eq.~(\ref{hb2}), we have%
\begin{equation}
H_{\text{quant}}(r)=2\ln\left[  1+\bar{n}\left(  1-\sqrt{\tau}\right)
\right]  ,\label{QHBquantumEPR2}%
\end{equation}
for $r\geq2\ln\left[  1+\bar{n}\left(  1-\sqrt{\tau}\right)  \right]  $. Then,
we have numerically checked that for $r<2\ln\left[  1+\bar{n}\left(
1-\sqrt{\tau}\right)  \right]  $, the QHB $H_{\text{quant}}(r)$ can become
infinite. This can be checked by computing the QHB $H_{\text{quant}}(r)$
directly from Eq.~(\ref{QHBdef}) and using the Gaussian formula for the
$s$-overlap (see Appendix~\ref{AppQHBdetail} for more details).

\section{Comparison and Quantum Advantage\label{sec5}}

In this section, we compare the classical and quantum strategies for
noninvasive sensing of loss, showing how the use of quantum correlations
enables us to outperform the classical benchmark achieved with coherent-state
inputs. For symmetric testing, we consider the difference (or gain)%
\begin{equation}
\Delta:=\bar{p}_{\text{coh}}-\bar{p}_{\text{quant}}^{\text{QCB}}\leq\bar
{p}_{\text{coh}}-\bar{p}_{\text{quant}},\label{gainEQ}%
\end{equation}
using the expressions in Eqs.~(\ref{CLASStransmitter})
and~(\ref{EPRtransmitter}). Its positivity is a sufficient condition for the
superiority of the quantum transmitter (while $\Delta\leq0$ corresponds to an
inconclusive comparison). In particular, for $\Delta$ close to $1/2$, we have
that $\bar{p}_{\text{quant}}\simeq0$ and $\bar{p}_{\text{coh}}\simeq1/2$,
which means that the quantum strategy allows for perfect sensing while the
classical strategy is equivalent to random guessing.

\begin{figure}[ptbh]
\vspace{0.2cm}
\par
\begin{center}
\includegraphics[width=0.45\textwidth] {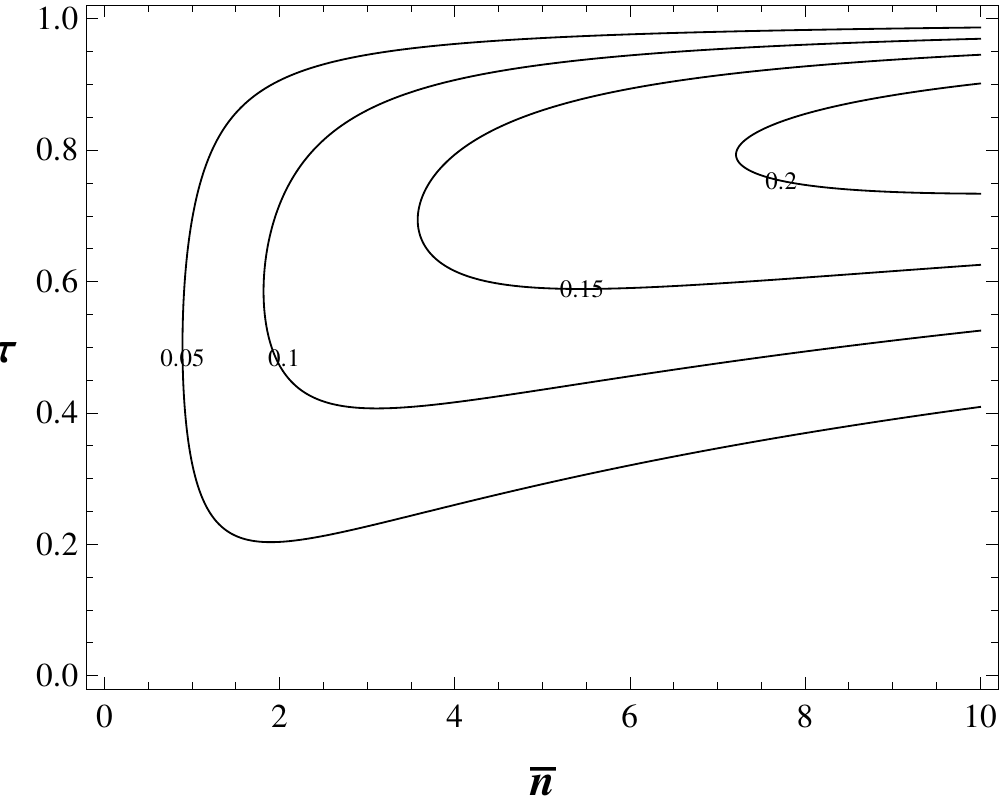}
\end{center}
\par
\vspace{-0.4cm}\caption{Comparison between the quantum transmitter (EPR
source) and the classical transmitter (coherent state) for single-copy probing
$M=1$. We contour-plot the gain $\Delta$ in the range $0\leq\tau<1$ and
$\bar{n}\leq10$. In most of the range, we have $\Delta>0$ proving the
superiority of the EPR\ transmitter, with better performances for $\tau
\simeq1$.}%
\label{EPRpic}%
\end{figure}

We start by considering a single probing of the sample, i.e., $M=1$. Then, we
plot $\Delta(\bar{n},\tau)$ considering the regime of small photon numbers
($\bar{n}\leq10$) and for $0\leq\tau<1$. As we can see from Fig.~\ref{EPRpic},
the quantum transmitter is better in most of the parameter plane, with very
good performances for $\tau$ close to $1$ (which corresponds to sensing an
almost transparent growth). To better explore this region we consider the
expansion for $\tau\simeq1$. By setting $\tau=1-\varepsilon$, we get the
first-order expansion%
\begin{equation}
\bar{p}_{\text{coh}}\simeq\frac{1}{2}\left(  1+\frac{\sqrt{\bar{n}}}%
{2}\varepsilon\right)  ,~~~~\bar{p}_{\text{quant}}^{\text{QCB}}\simeq\frac
{1}{2}\left(  1-\bar{n}\varepsilon\right)  ,
\end{equation}
so that
\begin{equation}
\Delta\simeq\left(  \frac{\sqrt{\bar{n}}+2\bar{n}}{4}\right)  \varepsilon,
\end{equation}
which is always positive. 

\begin{figure}[ptbh]
\vspace{0.3cm}
\par
\begin{center}
\includegraphics[width=0.45\textwidth] {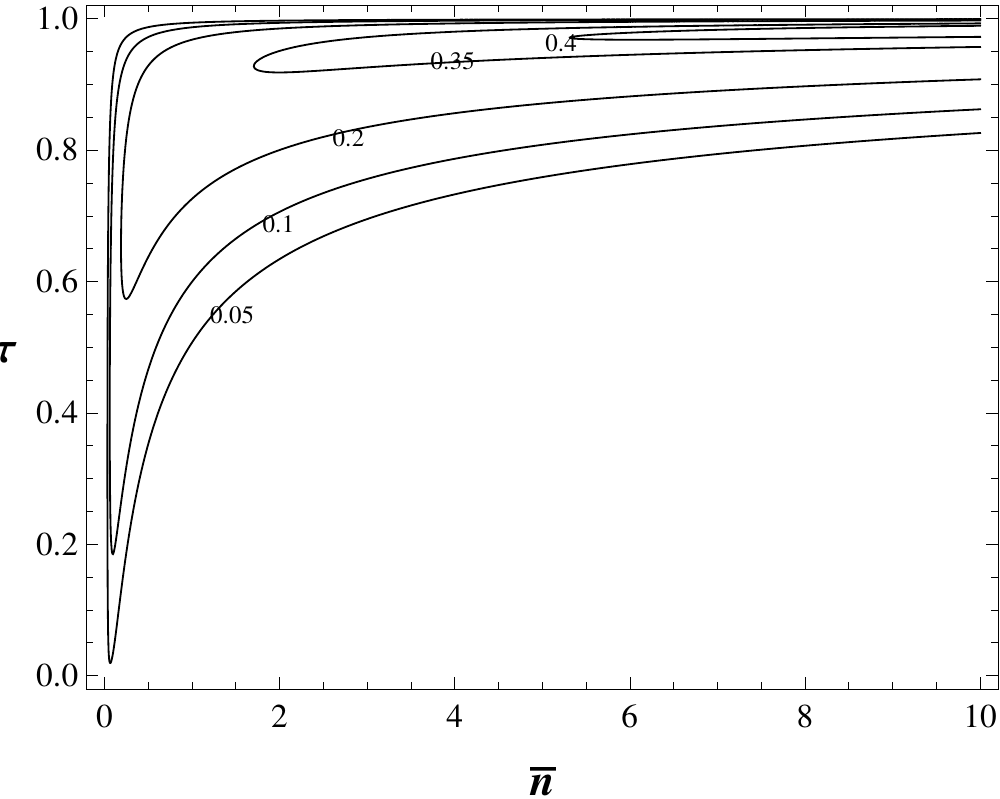}
\end{center}
\par
\vspace{-0.4cm}\caption{Comparison between the quantum transmitter (EPR
source) and the classical transmitter (coherent state) for multi-copy probing
$M=20$. We contour-plot the gain $\Delta$ in the range $0\leq\tau<1$ and
$\bar{n}\leq10$. In a good fraction of the range, we have $\Delta>0$ proving
the superiority of the EPR\ transmitter, with better performances occurring
for $\tau\simeq1$. Despite the fact they are not visible in the figure, for
$\bar{n}\simeq0$ we have $\Delta=0$ as expected. Similarly, we have $\Delta=0$
at exactly $\tau=1$.}%
\label{pic3}%
\end{figure}

We then analyze the multi-probing case where the samples are queried $M$ times
where we fix the mean number of photons per signal mode $\bar{n}$ (local
energy constraint). We then specify the gain $\Delta(M,\bar{n},\tau)$ for
$M=20$, and we plot the results in Fig.~\ref{pic3}. We see that the good
region where $\Delta$\ approaches the optimal value of $1/2$ is again for
transmissivities $\tau\simeq1$. 

\subsection{Asymptotic multi-copy behavior}

\begin{figure}[ptbh]
\vspace{+0.1cm}
\par
\begin{center}
\includegraphics[width=0.45\textwidth] {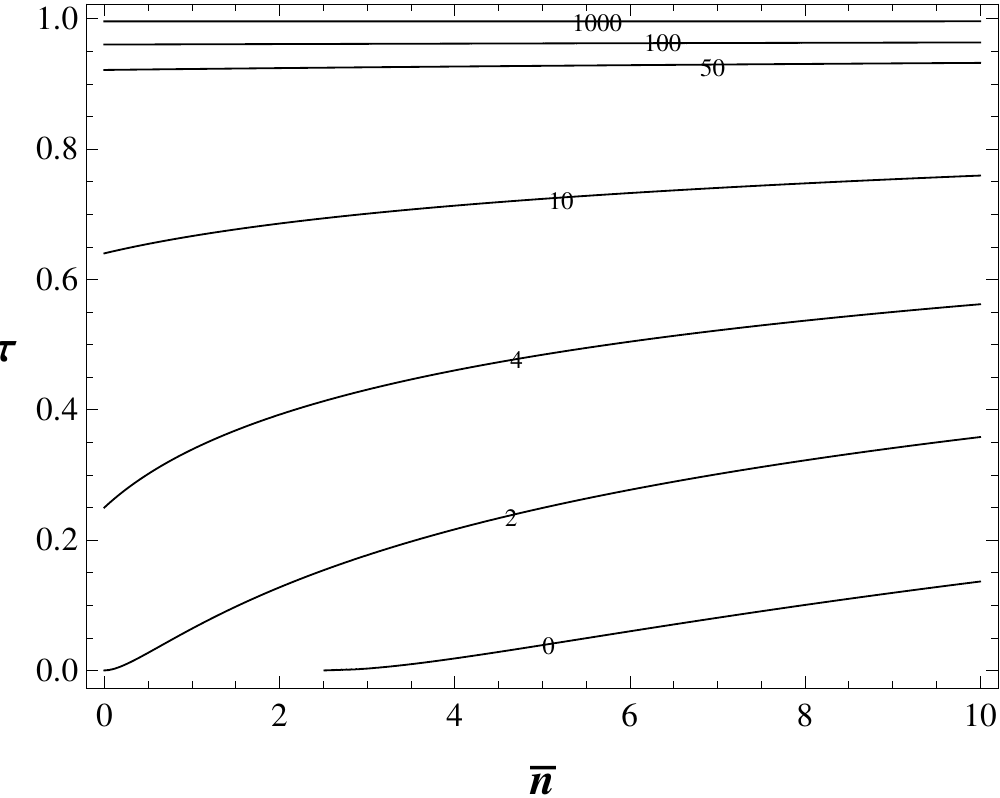}
\end{center}
\par
\vspace{-0.3cm}\caption{Plot of the asymptotic ratio $R=\kappa_{\text{quant}%
}/\kappa_{\text{coh}}$ in terms of transmissivity $\tau$ and mean number of
photons in the signal mode $\bar{n}$. Note that $R>1$ in almost all the plane
and\ $R$ becomes of the order of $10^{3}$ close to the region $\tau\simeq1$
(even though we have a discontinuity at exactly $\tau=1$, where $R$ must be
$1$).}%
\label{picRE}%
\end{figure}

Here we keep the local energy constraint, i.e., we fix the mean number of
photons per signal mode $\bar{n}$, and we compare the two transmitters in the
limit of a large number of copies $M\gg1$. In this limit the mean error
probability in the symmetric test is well approximated by the QCB, so that,
from Eqs.~(\ref{QCB1}) and~(\ref{EPRtransmitter}), we can write%
\begin{equation}
\bar{p}_{\text{coh}}\simeq\bar{p}_{\text{coh}}^{\text{QCB}}=\frac{1}%
{2}e^{-M\kappa_{\text{coh}}},
\end{equation}
where $\kappa_{\text{coh}}:=\bar{n}(1-\sqrt{\tau})^{2}$, and
\begin{equation}
\bar{p}_{\text{quant}}\simeq\bar{p}_{\text{quant}}^{\text{QCB}}=\frac{1}%
{2}e^{-M\kappa_{\text{quant}}},
\end{equation}
where $\kappa_{\text{quant}}:=2\ln[1+\bar{n}(1-\sqrt{\tau})]$. Thus, the
asymptotic gain can be measured by the ratio%
\begin{equation}
R=\frac{\kappa_{\text{quant}}}{\kappa_{\text{coh}}}~.\label{ra1}%
\end{equation}
It is clear that for $R>1$, the error probability of the EPR transmitter goes
to zero more rapidly than that of the coherent-state transmitter (while
$R\leq1$ corresponds to the opposite behavior). This ratio is shown in
Fig.~\ref{picRE}. We see that for high values of the transmissivity, the EPR
transmitter has an error exponent $\kappa_{\text{quant}}$ which becomes orders
of magnitude higher than the classical one $\kappa_{\text{coh}}$.

\subsection{Asymptotic multi-copy behavior: Asymmetric testing}

Assuming a local energy constraint we now compare the quantum and the
classical coherent transmitter from the point of view of asymmetric testing.
We consider the ratio between the two QHBs, i.e., for any control $r$ we
define%
\begin{equation}
R_{\text{QHB}}(r)=\frac{H_{\text{quant}}(r)}{H_{\text{coh}}(r)}~.\label{ra2}%
\end{equation}
According to Eqs.~(\ref{QHBcoherent}) and~(\ref{QHBquantumEPR}), we have%
\begin{equation}
H_{\text{coh}}(r)=\left\{
\begin{array}
[c]{l}%
r_{\text{coh}}~~\text{for~~}r\geq r_{\text{coh}}\\
\\
+\infty~~~\text{for~~}r<r_{\text{coh}}%
\end{array}
\right.
\end{equation}
and%
\begin{equation}
\left\{
\begin{array}
[c]{l}%
H_{\text{quant}}(r)=r_{\text{quant}}~~~\text{for~~}r\geq r_{\text{quant}}\\
\\
H_{\text{quant}}(r)\geq r_{\text{quant}}\text{~~~for~~}r<r_{\text{quant}}%
\end{array}
\right.
\end{equation}
where%
\begin{equation}
r_{\text{coh}}:=\bar{n}(1-\sqrt{\tau})^{2},~r_{\text{quant}}:=2\ln\left[
1+\bar{n}\left(  1-\sqrt{\tau}\right)  \right]  .
\end{equation}

If we assume good control on the false positives, i.e., $r\geq\max
\{r_{\text{coh}},r_{\text{quant}}\}$, then we find that the false negative
probability is well approximated by the QCB, i.e.,\ $p(0|1)\simeq\bar
{p}_{\text{QCB}}$ (see Sec.~\ref{SEConePoneM}). As a result, the ratio in
Eq.~(\ref{ra2}) asymptotically coincides with the previous ratio in
Eq.~(\ref{ra1}), i.e., $R_{\text{QHB}}(r)\simeq R$, and the same result shown
in Fig.~\ref{picRE}\ also applies to asymmetric testing.

\begin{figure}[ptbh]
\vspace{-0.0cm}
\par
\begin{center}
\includegraphics[width=0.45\textwidth] {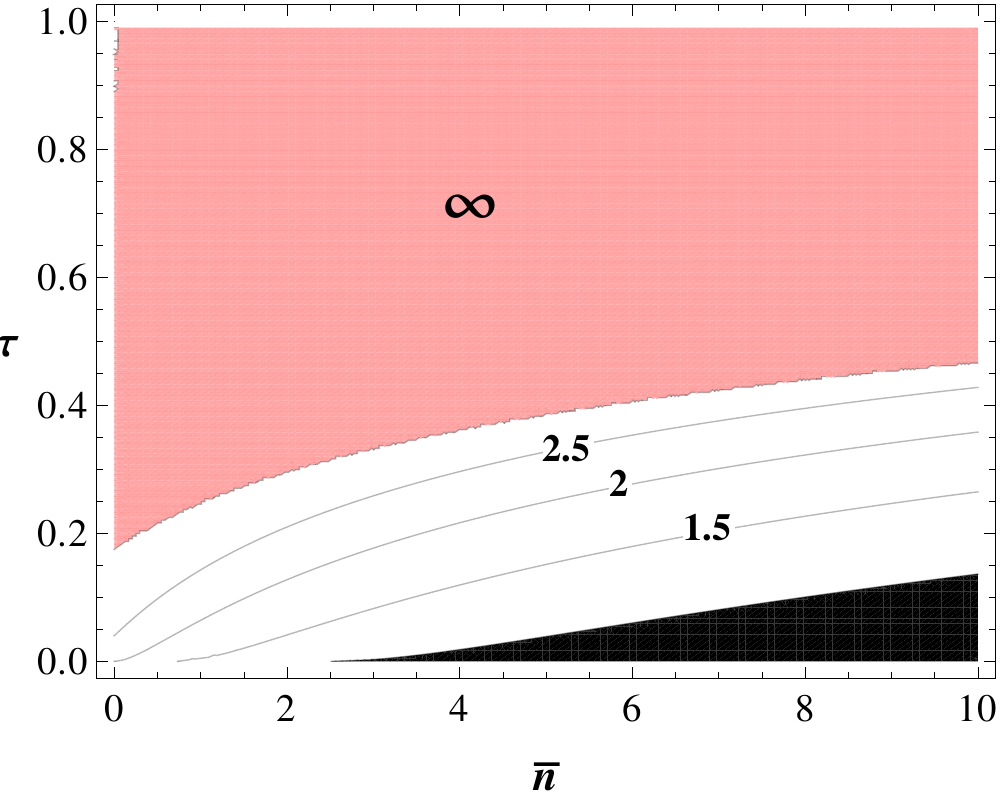}
\end{center}
\par
\vspace{-0.6cm}\caption{Contour-plot of the ratio $R_{\text{QHB}}(r)$ for
$r=r_{\text{coh}}<r_{\text{quant}}$ in the plane $(\bar{n},\tau)$. The black
area has to be ignored since it is not compatible with the condition
$r_{\text{coh}}<r_{\text{quant}}$. We can see that there is a region where the
ratio is finite and a wider one where it becomes infinite.}%
\label{QHBplot}%
\end{figure}

Let us now study what happens in the presence of moderate control of false
positives. Let us consider the case $r_{\text{coh}}<r_{\text{quant}}$ which
happens in a delimited region of the plane $(\bar{n},\tau)$ corresponding to
the non-black area in Fig.~\ref{QHBplot}. Then, we assume $r=r_{\text{coh}}$,
so that $H_{\text{coh}}(r)$ remains finite, while $H_{\text{quant}}(r)$ can
become infinite. As we see from Fig.~\ref{QHBplot}, there is a wide area where
$R_{\text{QHB}}(r)=+\infty$, meaning that the quantum EPR transmitter provides
a super-exponential decay for the false-negative probability, while it remains
exponential for the classical transmitter.

If we consider the opposite case $r=r_{\text{quant}}<r_{\text{coh}}$, which
may only occur in the black area of Fig.~\ref{QHBplot}, then we see that
$H_{\text{coh}}(r)$ becomes infinite while $H_{\text{quant}}(r)$ remains
finite. In other words, we have $R_{\text{QHB}}(r)=0$ in all the black region.
Finally, for $r<\min\{r_{\text{coh}},r_{\text{quant}}\}$ there are
indeterminate forms which do not allow us to provide a simple description of
the situation.

From these results it is clear that, for low photon numbers per mode ($\bar
{n}\lesssim10$) and high transmissivities (here $\tau\gtrsim0.5$), the quantum
EPR transmitter greatly outperforms the classical transmitter in the
asymmetric quantum discrimination of loss.

\subsection{Comparison under the global energy constraint}

\begin{figure}[ptbh]
\vspace{-0.0cm}
\par
\begin{center}
\includegraphics[width=0.45\textwidth] {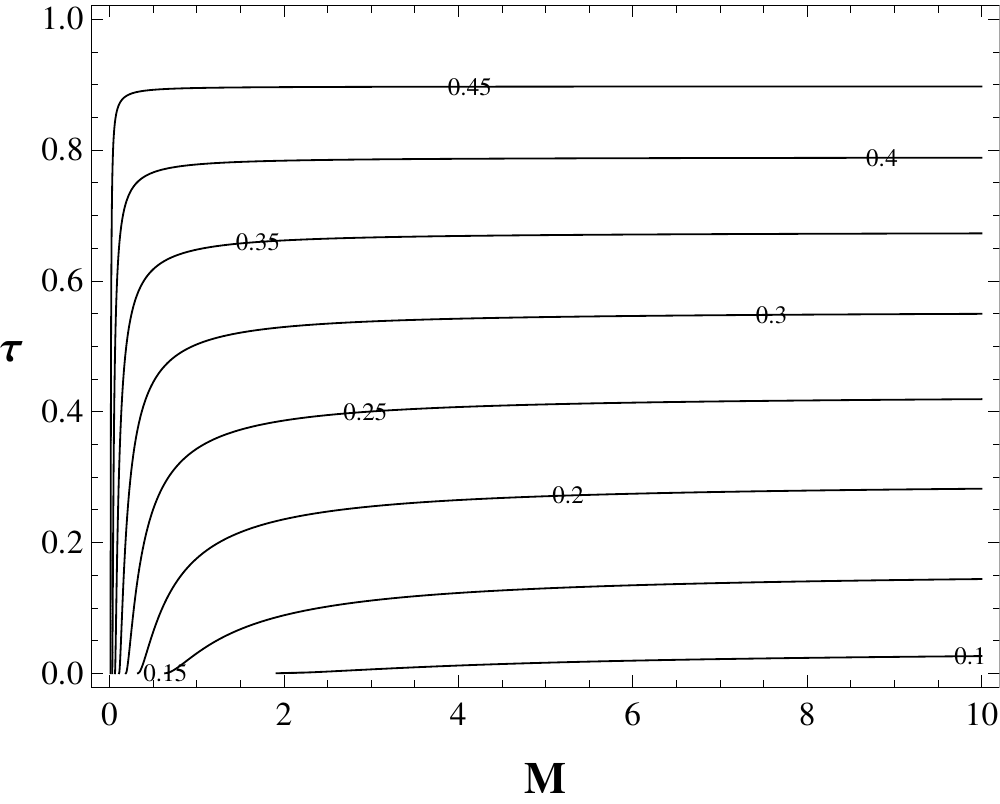}
\end{center}
\par
\vspace{-0.6cm}\caption{Behaviour of $\bar{p}_{\text{quant}}^{\text{QCB}}$ in
terms of $\tau$ and $M$ at fixed total energy $\bar{N}$. We see that at any
fixed transmissivity $\tau$, the QCB of the EPR transmitter is optimized by
increasing the number of copies $M$. This plot refers to $\bar{N}=1$ but the
behavior is generic for any $\bar{N}$.}%
\label{pic4}%
\end{figure}

In order to study the case of the global constraint we set $\bar{n}=\bar{N}/M$
in the QCB in Eq.~(\ref{EPRtransmitter}), so that it can expressed as $\bar
{p}_{\text{quant}}^{\text{QCB}}(\tau,\bar{N}/M,M)$. As we show in
Fig.~\ref{pic4}, the value of $\bar{p}_{\text{quant}}^{\text{QCB}}$ decreases
for increasing $M$, at fixed total energy $\bar{N}$ and transmissivity $\tau$.
In other words, it is better to use a large number of copies ($M\gg1$) of the
TMSV state with vanishingly small number of photons per copy ($\bar{n}\ll1$),
instead of a single energetic TMSV state with $\bar{N}$ mean photons
%
. Remarkably the asymptotic behavior is rapidly reached after a finite number
of copies, e.g., $M\simeq5$ for $\bar{N}=1$. 

\begin{figure}[ptbh]
\vspace{+0.1cm}
\par
\begin{center}
\includegraphics[width=0.45\textwidth] {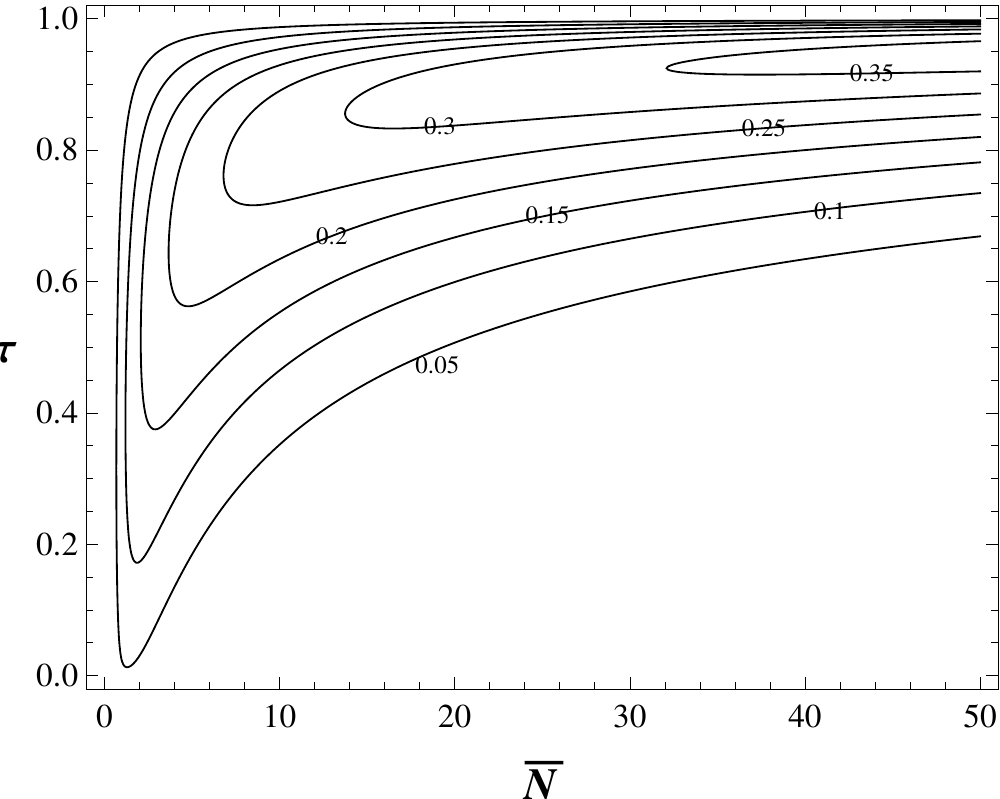}
\end{center}
\par
\vspace{-0.6cm}\caption{Contour-plot of the optimal gain $\Delta_{\text{opt}}$
as a function of $\tau$ and $\bar{N}$. Note that $\Delta_{\text{opt}}$
approaches $1/2$ in the top part of the plot.}%
\label{pic5}%
\end{figure}

For large $M$, we find the optimal asymptotic expression
\begin{equation}
\lim_{M\rightarrow+\infty}\bar{p}_{\text{quant}}^{\text{QCB}}(\tau,\bar
{N}/M,M)=\frac{1}{2}\exp\left[  -2\bar{N}(1-\sqrt{\tau})\right]  ,
\end{equation}
and we find the optimal gain is%
\begin{equation}
\Delta_{\text{opt}}:=\bar{p}_{\text{coh}}(\tau,\bar{N})-\frac{1}{2}\exp\left[
-2\bar{N}(1-\sqrt{\tau})\right]  .
\end{equation}
As shown in Fig.~\ref{pic5}, we can see that for relatively small numbers of
photons $\bar{N}\leq50$, globally irradiated over the sample the EPR
transmitter clearly outperforms the classical strategy, especially for high
transmissivities $\tau\simeq1$. In other words, the use of a quantum source
has non-trivial advantages for the noninvasive detection of small
concentrations. Luckily, we do not have to consider the limit of
$M\rightarrow\infty$ for approaching the optimal performance of the EPR
transmitter since, as we have already seen in Fig.~\ref{pic4}, this
performance is approximately reached with small finite $M$. Indeed, we have
numerically checked that $M=5$ already provides a good approximation of the
asymptotic behavior. 


\section{Biological growth and photodegradability\label{sec6}}

\subsection{Non-invasive detection of growth in biological samples}

As discussed in Sec.~\ref{sec2}, the concentration $c$ of species within a
sample can be connected with its transmissivity $\tau$ by the formula
$\tau(c)=10^{-\varepsilon lc}$, where $\varepsilon$\ is an absorption
coefficient and $l$ is the path length. Thus, the sample is a lossy channel
with concentration-dependent transmissivity $\tau(c)$ and the problem of loss
detection can be mapped into the discrimination between non-growth $(c=0)$ and
growth $(c>0)$ within the sample. For simplicity, in the following we set
$\varepsilon l=1$, so that $\tau(c)=10^{-c}$.

We then introduce a phenomenological model of bacterial/cell growth in the
sample, so that the concentration depends on time $t$ (abstract units)\ in a
typical exponential law
\begin{equation}
c(t)=c_{0}\left[  1-\exp(-gt)\right]  ~, \label{conce}%
\end{equation}
where $g$ is a saturation parameter and $c_{0}$ is the asymptotic
concentration (at infinite time). Using Eq.~(\ref{conce}) we can write $\tau$
as a function of time $t$ as follows%
\begin{equation}
\tau(t)=10^{-c_{0}\left[  1-\exp(-gt)\right]  }~. \label{conce2}%
\end{equation}

Now we can analyze how well we can distinguish between unit transmissivity (no
growth)\ and $\tau(t)<1$ (growth)\ at any specified time $t$. For this aim, we
replace $\tau(t)$ in the error probabilities associated with the classical and
quantum EPR transmitter. We consider the case of symmetric testing and we
assume a global energy constraint so that we fix the mean total number of
photons $\bar{N}$\ irradiated over the sample.

Mathematically, we then plug $\tau(t)$ into the formula of $\bar
{p}_{\text{coh}}(\tau,\bar{N})$ of Eq.~(\ref{CLASStransmitter}) for the
classical transmitter, so that we can write $\bar{p}_{\text{coh}}(t,\bar{N})$.
Similarly, for the quantum EPR transmitter, we plug $\tau(t)$ into the formula
of $\bar{p}_{\text{quant}}^{\text{QCB}}(\tau,\bar{n},M)$ of
Eq.~(\ref{EPRtransmitter}) where $\bar{n}M=\bar{N}$. In this case, we can
write $\bar{p}_{\text{quant}}^{\text{QCB}}(t,\bar{N},M)$ and study the two
extreme conditions of a single-mode EPR transmitter ($M=1$) and the broadband
EPR transmitter with $M\rightarrow+\infty$ (any other EPR transmitter with
arbitrary $M$ will have a performance between these two extremes).

\begin{figure}[ptbh]
\vspace{0.1cm}
\par
\begin{center}
\includegraphics[width=0.45\textwidth] {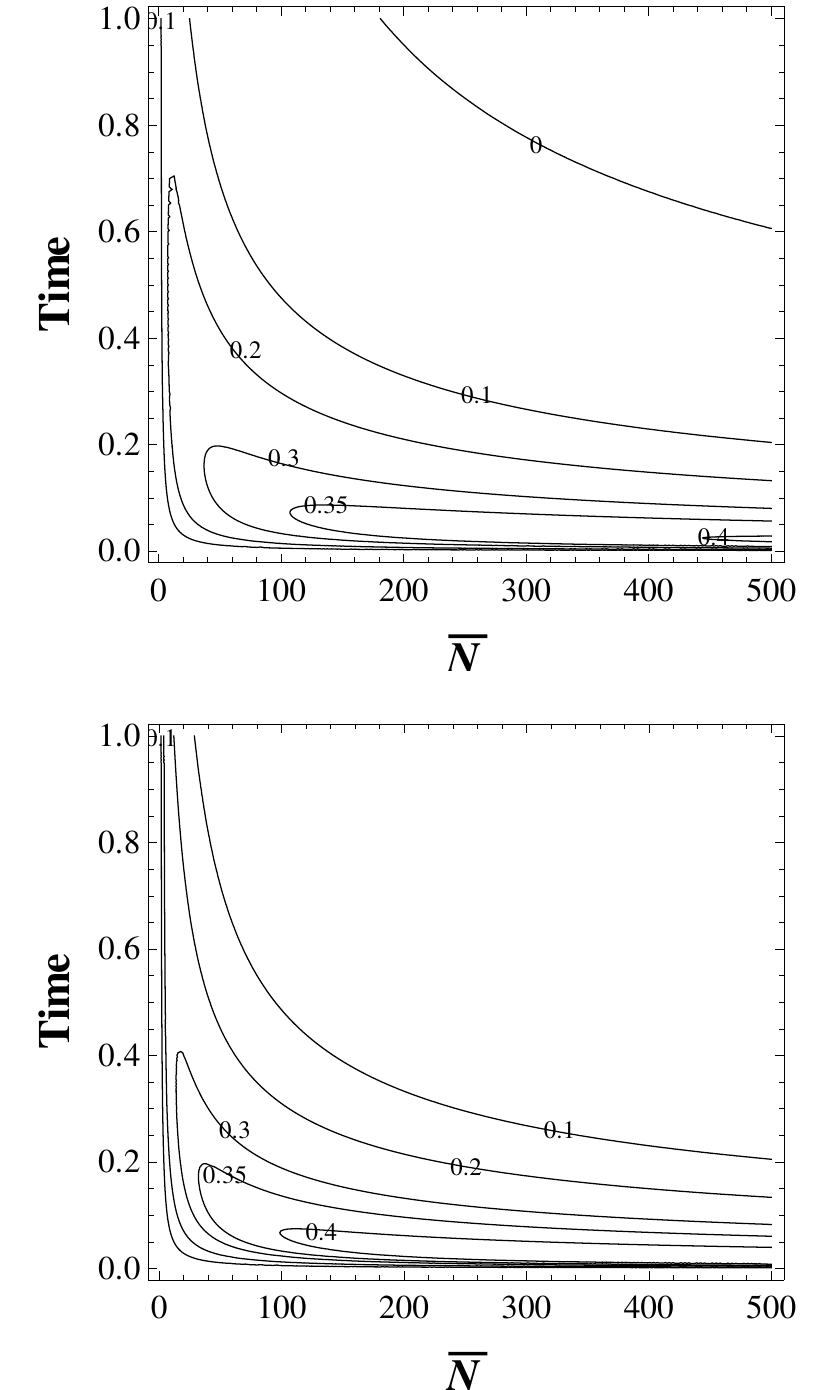}
\end{center}
\par
\vspace{-0.4cm}\caption{Contour-plot of the gains $\Delta_{1}(t,\bar{N})$ (top
panel)\ and $\Delta_{\text{opt}}(t,\bar{N})$ (bottom panel) versus mean total
energy $\bar{N}$ irradiated over the sample and time $t$ of detection. We can
see that the values approach $1/2$ at short timescales, corresponding to very
low concentrations. In both panels, we have chosen $c_{0}=1$ and $g=0.2$ for
the growth model of Eq.~(\ref{conce}).}%
\label{growthplot}%
\end{figure}

As before we can make the comparison by using the gain $\Delta=\bar
{p}_{\text{coh}}-\bar{p}_{\text{quant}}^{\text{QCB}}$ whose maximal value is
$1/2$. Specifically, we consider the gain given by the single mode EPR
transmitter
\begin{equation}
\Delta_{1}(t,\bar{N})=\bar{p}_{\text{coh}}(t,\bar{N})-\bar{p}_{\text{quant}%
}^{\text{QCB}}(t,\bar{N},1),
\end{equation}
and the optimal gain given by the broadband EPR transmitter, i.e.,%
\begin{equation}
\Delta_{\text{opt}}(t,\bar{N})=\bar{p}_{\text{coh}}(t,\bar{N})-\bar
{p}_{\text{quant}}^{\text{QCB}}(t,\bar{N},+\infty).
\end{equation}
We compare the performances of the classical and quantum transmitters plotting
$\Delta_{1}(t,\bar{N})$ and $\Delta_{\text{opt}}(t,\bar{N})$ in
Fig.~\ref{growthplot}.

\begin{figure}[ptbh]
\vspace{0.2cm}
\par
\begin{center}
\includegraphics[width=0.45\textwidth] {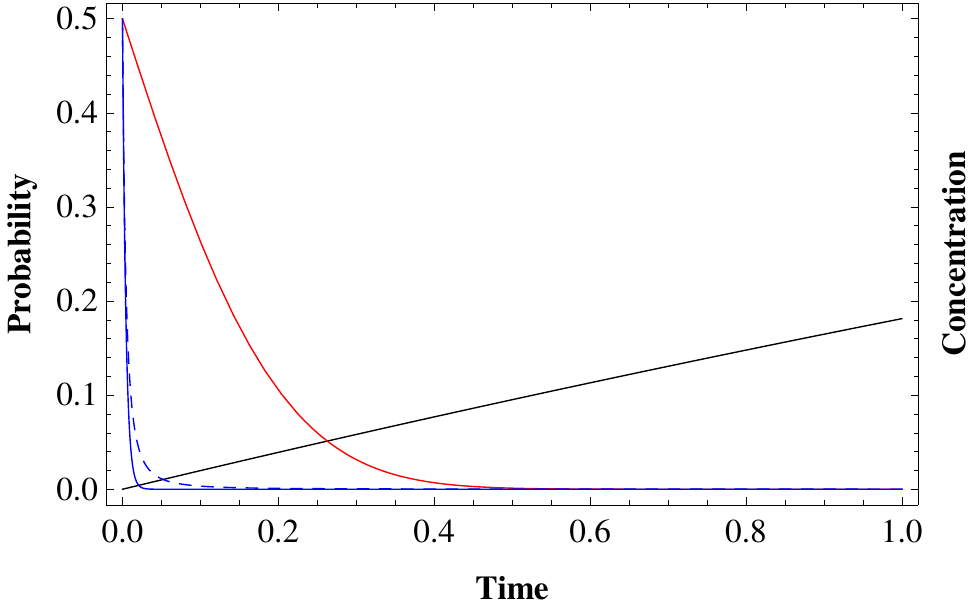}
\end{center}
\par
\vspace{-0.6cm}\caption{Error probabilities of the various transmitters
(operated at $\bar{N}=500$ photons) as a function of time $t$ (abstract
units). The red curve refers to the probability of the classical coherent
transmitter $\bar{p}_{\text{coh}}$. The blue curves refer to the quantum EPR
transmitter $\bar{p}_{\text{quant}}^{\text{QCB}}$ for single-mode probing
$M=1$ (dashed curve) and broadband probing $M\rightarrow+\infty$ (solid
curve). We also show the corresponding behaviour of the concentration (solid
black curve) which increases in time. Here we have chosen $c_{0}=1$ and
$g=0.2$ for the growth model of Eq.~(\ref{conce}).}%
\label{Timeplot}%
\end{figure}

As we can see from Fig.~\ref{growthplot}, the EPR transmitter outperforms the
classical strategy at short times, i.e., at low concentrations, when the mean
total number of photons $\bar{N}$ is restricted to relatively small values.
This means that, in this non-destructive regime, the EPR\ transmitter is able
to provide a much faster detection of bacterial/cell growth in the sample.
This is also evident from Fig.~\ref{Timeplot}, where we explicitly compare the
performances of the transmitters at $\bar{N}=500$ photons. As we can see, the
quantum transmitter allows us to detect the presence or absence of a growth in
extremely short times ($<0.05$ time units in the figure), while we need to
wait longer times (at least $0.4$ time units) for obtaining the same
performance by means of a classical transmitter.

\subsection{Model combining growth and photo-degradability}

We now combine the previous exponential growth model with a model of
photo-degradability. We assume that the concentration decays for increasing
photon number $\bar{N}$. More precisely, we assume that the concentration
decay by a factor $\exp(-\gamma\bar{N}t)$ where $\gamma$ is a degradability
parameter, $\bar{N}$ is the mean number of photons irradiated over the sample
(constantly per unit of time), and $t$ is the time of the readout. We assume
that at time $t$ the sample has been already irradiated with $\bar{N}t$ mean
photons, and the duration of the readout is equal to one time unit. Combining
this process with the growth model, we have
\begin{equation}
c(t)=c_{0}\left[  1-\exp(-gt)\right]  \exp(-\gamma\bar{N}t)~, \label{combb}%
\end{equation}
and a corresponding model for the transmissivity $\tau(t)=10^{-c(t)}$
according to the Beer-Lambert law.

\begin{figure}[ptbh]
\vspace{0.15cm}
\par
\begin{center}
\includegraphics[width=0.45\textwidth] {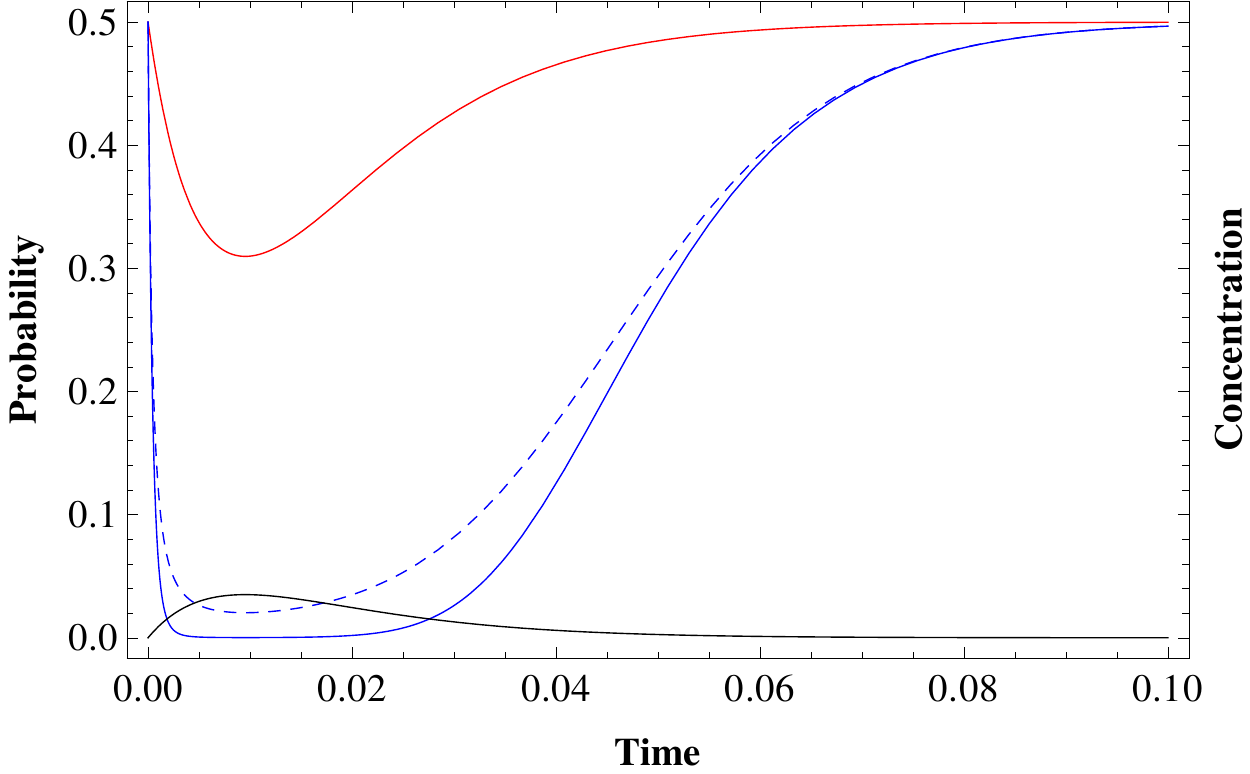}
\end{center}
\par
\vspace{-0.3cm}\caption{Error probabilities of the various transmitters as a
function of time $t$ (abstract units) and assuming $\bar{N}=100$ mean photons
irradiated per time unit. The red curve refers to the error probability of the
classical coherent transmitter $\bar{p}_{\text{coh}}$. The blue curves refer
to the quantum EPR transmitter $\bar{p}_{\text{quant}}^{\text{QCB}}$ for
single-mode probing $M=1$ (dashed curve) and broadband probing $M\rightarrow
+\infty$ (solid curve). We also show the corresponding behaviour of the
concentration (solid black curve). We have chosen $c_{0}=\gamma=1$, $g=10$ in
the photo-degradable model of Eq.~(\ref{combb}).}%
\label{TimeDEG}%
\end{figure}

For simplicity, here we set $c_{0}=\gamma=1$, $g=10$, and $\bar{N}=100$. In
Fig.~\ref{TimeDEG} we show the behavior of the concentration versus time and
the corresponding performances of the quantum and classical transmitters. The
plot shows that the quantum transmitter clearly outperforms the classical
benchmark in terms of fast detection times. As a matter of fact, in the regime
of parameters investigated, the growth is present only in a small time window
(before the sample is degraded). Within this window, the quantum transmitter
is able to detect bacterial growth with certainty, while the classical
transmitter has a non-trivial error probability ($\gtrsim0.3$%
).

\section{Fragile data storage\label{sec7}}

A practical scenario where the quantum detection of loss is important is that
of quantum reading~\cite{Qread}, where the aim is to boost the retrieval of
classical data from an optical disk by exploiting quantum entanglement at low
photon numbers. A basic setup for quantum reading consists of a series of
cells, each one encoding a bit of information. This is physically done by
picking two different reflectivities, $r_{0}$ and $r_{1}$, each encoding a
different bit value. A target cell is then read by shining optical modes on it
(generated by the transmitter) and detecting their reflection back to a
receiver. The use of an entangled source is known to retrieve more information
than any classical source (e.g., based on a mixture of coherent states).

Here we consider the same basic model but in transmission, i.e., assuming that
the bit-values are encoded in two different transmissivities, $\tau_{0}$ and
$\tau_{1}$, and with the cells positioned between the transmitter and the
receiver. This is completely equivalent from a mathematical point of view
(since $\tau+r=1$). Similarly to Ref.~\cite{Qread9b}, we show that the
advantage given by a quantum EPR transmitter can be made extreme by
introducing a suitable photodegradable model for the memory. In fact, we can
assume a specific saturation behavior for the memory cells in such a way that
their transmissivities tend to coincide if we increase the amount of energy
adopted for their readout. Assuming this model, we find wide regions of
parameters where the EPR transmitter is able to retrieve the maximum value of
$1$ bit per cell, while the classical coherent transmitter decodes $\simeq0$
bits per cell. As we discuss below, this striking difference can be exploited
as a cryptographic technique.

As before we assume that $\tau_{0}=1$ and $\tau_{1}:=\tau<1$. We can introduce
a simple saturation effect by imposing that the lower transmissivity $\tau$
tends to $1$ for increasing photon numbers. This may be realized by imposing
the exponential law%
\begin{equation}
\tau=1-\theta_{1}\exp(-\theta_{2}\bar{N}), \label{satu}%
\end{equation}
where $\bar{N}$ is the mean total number of photons employed in each probing,
while $\theta_{1}$ and $\theta_{2}$ are parameters of the phenomenological
model.
More specifically, parameter $\theta_{1}$\ provides the value at zero energy
(which is $1-\theta_{1}$), and parameter $\theta_{2}$\ gives the speed of
convergence to $1$.

In order to evaluate the effects of this saturation behavior, we have to
combine the law of Eq.~(\ref{satu}) with the energy-dependent performances of
the quantum and classical transmitters. First of all, since we are considering
the readout of a memory cell, we connect the error probability in the channel
discrimination $\bar{p}$ with the amount of information retrieved $I$. This
connection is given by the formula%
\begin{equation}
I(\bar{p})=1-H(\bar{p}),
\end{equation}
where $H(\bar{p}):=-\bar{p}\log_{2}\bar{p}-(1-\bar{p})\log_{2}(1-\bar{p})$ is
the binary Shannon entropy. Thus, for the coherent transmitter, we have
$I_{\text{coh}}:=I(\bar{p}_{\text{coh}})$ where $\bar{p}_{\text{coh}}%
(\tau,\bar{N})$ is given in Eq.~(\ref{CLASStransmitter}). For the quantum EPR
transmitter, we have $I_{\text{quant}}:=I(\bar{p}_{\text{quant}}^{\text{QCB}%
})\leq I(\bar{p}_{\text{quant}})$, where the QCB $\bar{p}_{\text{quant}%
}^{\text{QCB}}(\tau,\bar{n},M)$\ is given in Eq.~(\ref{EPRtransmitter}) and
$\bar{n}M=\bar{N}$. Thus, for any fixed choice of the parameters $\theta_{1}$
and $\theta_{2}$, we can replace $\tau(\bar{N})$ in the previous information
quantities, so to have $I_{\text{coh}}=I_{\text{coh}}(\bar{N})$ and
$I_{\text{quant}}=I_{\text{quant}}(\bar{N},M)$.

\begin{figure*}[ptbh]
\vspace{-0.5cm}
\par
\begin{center}
\includegraphics[width=0.95\textwidth] {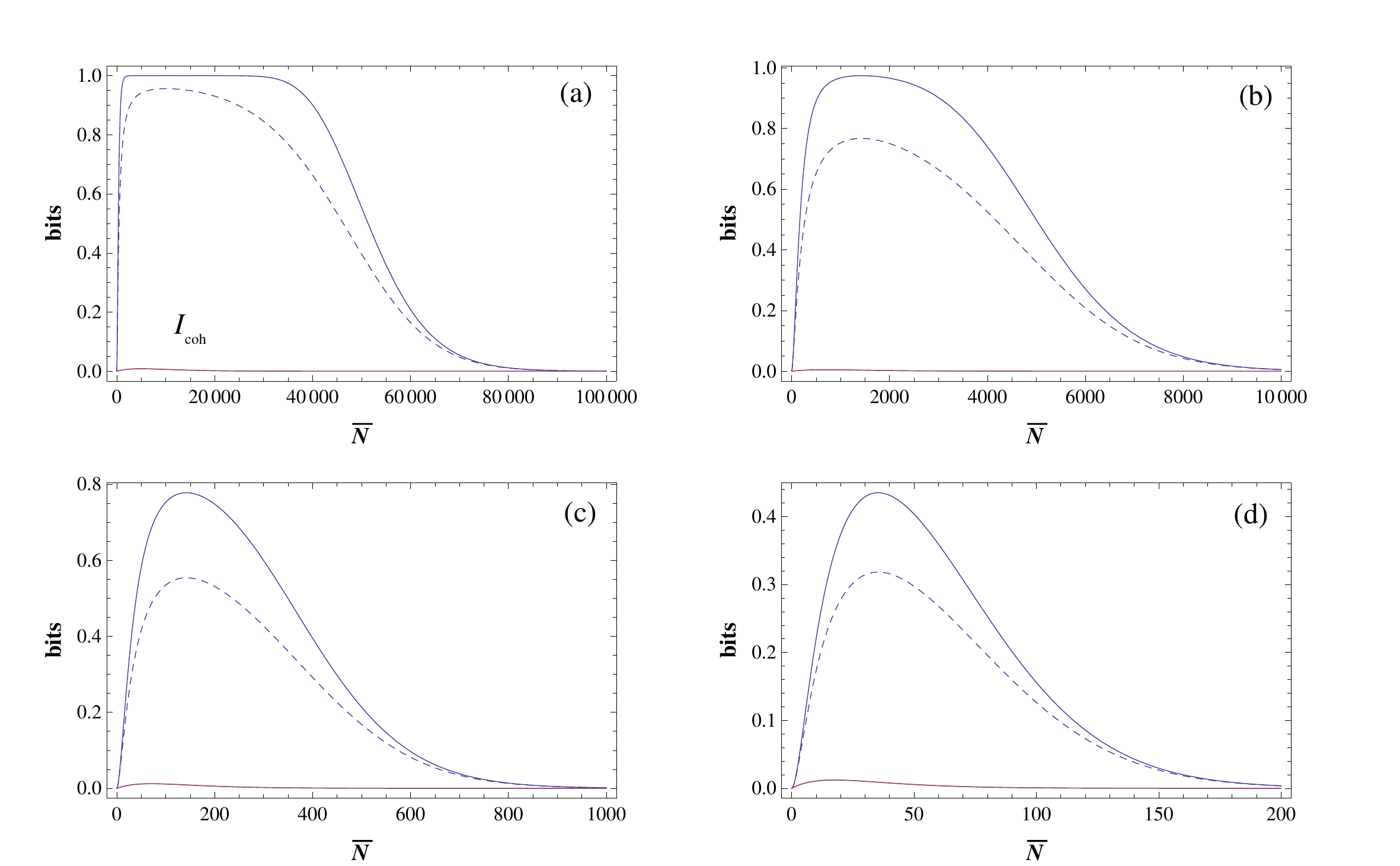}
\end{center}
\par
\vspace{-0.3cm}\caption{Number of bits per cell which are retrieved by
irradiating $\bar{N}$ mean total number of photons per cell. In each panel,
the red\ curve close to zero is the performance of the classical coherent
transmitter $I_{\text{coh}}(\bar{N})$. The dashed blue curve is the
performance of a single-mode EPR transmitter $I_{\text{quant}}(\bar{N},M=1)$,
while the solid blue curve is the performance of a broadband EPR transmitter
$I_{\text{quant}}(\bar{N},M\rightarrow+\infty)$. Any EPR transmitter at fixed
energy $\bar{N}$ and arbitrary number of modes per cell $M$ has a performance
between the dashed and the solid blue curves. The panels refer to various
choices for $\theta_{1}$ and $\theta_{2}$ in the exponential law of
Eq.~(\ref{satu}). We consider: (a)$~(\theta_{1},\theta_{2})=(5\times
10^{-3},10^{-4})$; (b)$~(\theta_{1},\theta_{2})=(10^{-2},7\times10^{-4})$;
(c)$~(\theta_{1},\theta_{2})=(5\times10^{-2},7\times10^{-3})$; and
(d)$~(\theta_{1},\theta_{2})=(10^{-1},28\times10^{-3})$.}%
\label{ExtremeREAD}%
\end{figure*}

At fixed values of the energy $\bar{N}$, we then compare the number of bits
retrieved by the classical transmitter $I_{\text{coh}}(\bar{N})$ with those
retrieved by the quantum EPR transmitter $I_{\text{quant}}(\bar{N},M)$ for
$M=1$ (i.e., a single energetic mode with $\bar{N}$ mean photons) and for
$M\rightarrow+\infty$ (i.e., an infinite number of modes with vanishing mean
photons $\bar{N}/M$). The performance of the quantum transmitter for arbitrary
$M$ will be bounded by these two extremal curves. This comparison is shown in
Fig.~\ref{ExtremeREAD}\ where we assume several values for the parameters
$\theta_{1}$ and $\theta_{2}$. From the previous figure it is clear that we
can consider photodegradable models such that the classical transmitter is not
able to retrieve any information, while the EPR\ transmitter can retrieve
almost all the information in a specific range of energies, depending on the
$\theta$'s.

This extreme situation could be exploited for cryptographic tasks. In
particular, an optical memory could be purposely constructed to be
photo-degradable in such a way as to hide its encoded classical data from any
standard optical drives based on the use of coherent (or thermal) light. Only
an advanced laboratory able to engineer a quantum entangled source in the
correct window of energy will be able to read out the stored confidential
data. From this point of view, quantum reading can provide a potential
technological layer of security based on the fact that the generation of
entanglement and other non-classical features is only possible in more
advanced labs of quantum optics. Furthermore, the range of energy to be used
must also be very well-tailored depending on the specific parameters
$\theta_{1}$ and $\theta_{2}$ employed during data storage which means that
even an eavesdropper with an advanced laboratory is likely to destroy the
data. These concepts are clearly preliminary but the basic ideas could further
be developed into potential practical applications.

\section{Conclusions\label{sec8}}

Partly building on previous analyses (e.g., Ref.~\cite{Qread}), we have
considered the quantum discrimination of bosonic loss and extended results
from symmetric to asymmetric testing. In particular, we have derived the
quantum Hoeffding bound associated with the discrimination of loss by means of
coherent states (classical benchmark/transmitter) and TMSV states (quantum
EPR\ transmitter). Then, we have applied symmetric testing to the study of
fragile systems. In fact, we have shown how the use of quantum entanglement
greatly outperforms classical strategies (based on coherent-state
transmitters) for the low-energy detection of small concentrations in samples.
Assuming models of bacterial growth and photo-degradability, we have shown
parameter regimes where quantum entanglement guarantees a very fast detection
of growth, while coherent-state transmitters are very slow or even unable to
detect the presence of bacteria.

Similar results are shown for the readout of photo-degradable optical
memories. Future steps involve the study of the quantum advantage by
considering other phenomenological models of bacterial growth and
photo-degradability. In the long run, these principles can be exploited to
design more advanced types of biological instrumentations, such as noninvasive
quantum-enhanced spectrophotometers for concentration detection and
measurement. Because of the absence of degradation, fragile samples could also
be re-used or continuously probed in a real-time fashion.

\section{Acknowledgments}

This work has been sponsored by the European Commission under the Marie
Sklodowska-Curie Fellowship Progamme (EC\ grant agreement No 745727), the
EPSRC and the Leverhulme Trust (`qBIO' fellowship). G.S. would like to thank
Seth Lloyd for feedback.

\appendix

\section{Formulas for Gaussian states\label{Gauss_APP}}

For the sake of completeness we provide the formula for the $s$-overlap
between two Gaussian states, originally proven in Ref.~\cite{MinkoPRA}. This
formula enables us to compute the QCB in Eq.~(\ref{QCBdef}) and the QHB in
Eq.~(\ref{QHBdef}). Here we specify this formula for the case of two-mode
Gaussian states.

Given a pair of two-mode Gaussian states $\sigma_{0}$ and $\sigma_{1}$, their
$s$-overlap can be computed in terms of their first- and second-order
statistical moments, i.e., their mean values, $\mathbf{\bar{x}}_{0}$ and
$\mathbf{\bar{x}}_{1}$, and their CMs, $\mathbf{V}_{0}$ and $\mathbf{V}_{1}$.
In particular, the CMs can be decomposed as
\begin{align}
\mathbf{V}_{0}  &  =\mathbf{S}_{0}\left(
\begin{array}
[c]{cc}%
\nu_{-}^{0}\mathbf{I} & \mathbf{0}\\
\mathbf{0} & \nu_{+}^{0}\mathbf{I}%
\end{array}
\right)  \mathbf{S}_{0}^{T},\label{v0app}\\
\mathbf{V}_{1}  &  =\mathbf{S}_{1}\left(
\begin{array}
[c]{cc}%
\nu_{-}^{1}\mathbf{I} & \mathbf{0}\\
\mathbf{0} & \nu_{+}^{1}\mathbf{I}%
\end{array}
\right)  \mathbf{S}_{1}^{T}, \label{v1app}%
\end{align}
where $\{\nu_{-}^{0},\nu_{+}^{0}\}$ and $\{\nu_{-}^{1},\nu_{+}^{1}\}$ are
their symplectic spectra, and $\mathbf{S}_{0}$ and $\mathbf{S}_{1}$ are
symplectic matrices~\cite{RMP}.

Let us introduce the following real functions depending on the parameter
$0<s\leq1$
\begin{align}
G_{s}(x) &  :=\frac{2^{s}}{(x+1)^{s}-(x-1)^{s}},\label{Gfunc}\\
\Lambda_{s}(x) &  :=\frac{(x+1)^{s}+(x-1)^{s}}{(x+1)^{s}-(x-1)^{s}%
}.\label{Lfunc}%
\end{align}
Then, the $s$-overlap can be written as%
\begin{equation}
C_{s}=\frac{\Pi_{s}}{\sqrt{\det\boldsymbol{\Sigma}_{s}}}\exp\left[
-\frac{\mathbf{d}^{T}\boldsymbol{\Sigma}_{s}^{-1}\mathbf{d}}{2}\right]
,\label{sOVERLAP}%
\end{equation}
where $\mathbf{d}:=\mathbf{\bar{x}}_{0}-\mathbf{\bar{x}}_{1}$,%
\begin{equation}
\Pi_{s}:=4G_{s}(\nu_{+}^{0})G_{s}(\nu_{-}^{0})G_{1-s}(\nu_{+}^{1})G_{1-s}%
(\nu_{-}^{1})~,
\end{equation}
and%
\begin{align}
\boldsymbol{\Sigma}_{s}  & :=\mathbf{S}_{0}\left(
\begin{array}
[c]{cc}%
\Lambda_{s}(\nu_{-}^{0})\mathbf{I} & \mathbf{0}\\
\mathbf{0} & \Lambda_{s}(\nu_{+}^{0})\mathbf{I}%
\end{array}
\right)  \mathbf{S}_{0}^{T}\nonumber\\
& +\mathbf{S}_{1}\left(
\begin{array}
[c]{cc}%
\Lambda_{1-s}(\nu_{-}^{1})\mathbf{I} & \mathbf{0}\\
\mathbf{0} & \Lambda_{1-s}(\nu_{+}^{1})\mathbf{I}%
\end{array}
\right)  \mathbf{S}_{1}^{T}.\label{SIGMA}%
\end{align}

Note that for computing the $s$-overlap it is crucial to derive the full
symplectic decompositions of Eqs.~(\ref{v0app}) and~(\ref{v1app}), not only
the symplectic spectra. However, this is an easy task for CMs of the form%
\begin{equation}
\mathbf{V}=\left(
\begin{array}
[c]{cc}%
a\mathbf{I} & c\mathbf{Z}\\
c\mathbf{Z} & b\mathbf{I}%
\end{array}
\right)  .\label{formEASY}%
\end{equation}
In this case we can write~\cite{RMP}
\begin{equation}
\nu_{\pm}=\frac{\sqrt{y}\pm(b-a)}{2},~~~~y:=(a+b)^{2}-4c^{2}%
,\label{spectrumEASY}%
\end{equation}
and%
\begin{equation}
\mathbf{S}=\left(
\begin{array}
[c]{cc}%
\omega_{+}\mathbf{I} & \omega_{-}\mathbf{Z}\\
\omega_{-}\mathbf{Z} & \omega_{+}\mathbf{I}%
\end{array}
\right)  ,~~~~\omega_{\pm}:=\sqrt{\frac{a+b\pm\sqrt{y}}{2\sqrt{y}}%
.}\label{esseEASY}%
\end{equation}

Finally, we also provide the formula for the fidelity between two multi-mode
Gaussian states, one pure with statistical moments $\{\mathbf{x}%
_{0},\mathbf{V}_{0}\}$ and one mixed with statistical moments $\{\mathbf{x}%
_{1},\mathbf{V}_{1}\}$. This is given by~\cite{Spedalieri13}%
\begin{equation}
F=\frac{2^{n}}{\sqrt{\det(\mathbf{V}_{0}+\mathbf{V}_{1})}}\exp\left[
-\frac{\mathbf{d}^{T}\left(  \mathbf{V}_{0}+\mathbf{V}_{1}\right)
^{-1}\mathbf{d}}{2}\right]  , \label{Fid_FormulaNEW}%
\end{equation}
where $\mathbf{d}:=\mathbf{\bar{x}}_{0}-\mathbf{\bar{x}}_{1}$. We use this
formula in Eqs.~(\ref{fidcoh}) and~(\ref{Fgauss2}) of the main text.

\section{Single-mode Gaussian channels and lossy channels\label{APP_lossy}}

Let us consider a single bosonic mode in a Gaussian state $\rho$, with mean
value $\mathbf{\bar{x}}$ and covariance matrix (CM) $\mathbf{V}$. The action
of a Gaussian channel on this state $\rho\rightarrow\rho^{\prime}%
=\mathcal{E}(\rho)$ can be easily expressed in terms of simple transformations
on its statistical moments. In particular, we have~\cite{RMP}%
\begin{align}
\mathbf{\bar{x}}  &  \rightarrow\mathbf{\bar{x}}^{\prime}=\mathbf{K\bar
{x}+d~,}\label{s1}\\
\mathbf{V}  &  \rightarrow\mathbf{V}^{\prime}=\mathbf{KVK}^{T}+\mathbf{N~,}
\label{s2}%
\end{align}
where $\mathbf{d}$ is a displacement vector, $\mathbf{K}$ a transmission
matrix, and $\mathbf{N}$ a noise matrix (satisfying suitable bona-fide
conditions~\cite{RMP}).

A lossy channel is a specific Gaussian channel described by $\mathbf{d=0}$
and
\begin{equation}
\mathbf{K=}\sqrt{\tau}~\mathbf{I},~~~~\mathbf{N}=(1-\tau)\mathbf{I},
\end{equation}
with $\tau\in\lbrack0,1]$ being the transmissivity. From Eqs.~(\ref{s1})
and~(\ref{s2}), we see that the output statistical moments are given by
$\mathbf{\bar{x}}^{\prime}=\sqrt{\tau}~\mathbf{\bar{x}}$ and $\mathbf{V}%
^{\prime}=\tau\mathbf{V}+(1-\tau)\mathbf{I}$.

\section{Computation of the CM in Eq.~(\ref{CM1out})\label{APP_CM}}

To compute the CM of the output state $\rho_{1}$, we dilate the lossy channel
into a beam splitter (with transmissivity $\tau$), mixing the signal mode $S$
with a vacuum mode $v$. Thus, we have a Gaussian unitary transformation from
the input state $\rho_{\text{in}}=\left\vert 0\right\rangle _{v}\left\langle
0\right\vert \otimes\left\vert \mu\right\rangle _{SR}\left\langle
\mu\right\vert $ of modes $(v,S,R)$ into the output state $\rho_{\text{out}}$
of modes $(v^{\prime},S^{\prime},R)$, i.e.,%
\begin{equation}
\rho_{\text{out}}=[\hat{U}_{vS}(\tau)\otimes\hat{I}_{R}]\rho_{\text{in}}%
[\hat{U}_{vS}(\tau)\otimes\hat{I}_{R}]^{\dagger},
\end{equation}
where $\hat{U}_{vS}(\tau)$ is the beam-splitter unitary~\cite{RMP} applied to
modes $v$ and $S$, while the reference mode $R$ is subject to the identity. In
terms of CMs, we have%
\begin{equation}
\mathbf{V}_{\text{out}}=[\mathbf{B}_{vS}(\tau)\oplus\mathbf{I}_{R}%
]\mathbf{V}_{\text{in}}[\mathbf{B}_{vS}(\tau)\oplus\mathbf{I}_{R}]^{T},
\end{equation}
where $\mathbf{V}_{\text{in}}=\mathbf{I}_{v}\oplus\mathbf{V}_{SR}(\mu)$ and%
\begin{equation}
\mathbf{B}_{vS}(\tau)=\left(
\begin{array}
[c]{cc}%
\sqrt{\tau}\mathbf{I} & \sqrt{1-\tau}\mathbf{I}\\
-\sqrt{1-\tau}\mathbf{I} & \sqrt{\tau}\mathbf{I}%
\end{array}
\right)  ,
\end{equation}
is the symplectic transformation of the beam splitter. After simple algebra,
we find an output CM of the form%
\begin{equation}
\mathbf{V}_{\text{out}}=\left(
\begin{array}
[c]{cc}%
\mathbf{W} & \mathbf{C}\\
\mathbf{C}^{T} & \mathbf{V}_{1}(\mu,\tau)
\end{array}
\right)  ,
\end{equation}
where the blocks $\mathbf{W}$ and $\mathbf{C}$ are to be traced out, while
$\mathbf{V}_{1}(\mu,\tau)$ is given in Eq.~(\ref{CM1out}).

\section{QHB for the quantum transmitter\label{AppQHBdetail}}

Here we compute the QHB\ for the quantum EPR transmitter $H_{\text{quant}}(r)$
directly from the Gaussian formula for the $s$-overlap given in
Appendix.~\ref{Gauss_APP}. Since the two output states generated by the EPR
transmitter are zero-mean, we can write the $s$-overlap as
\begin{equation}
C_{s}=\frac{\Pi_{s}}{\sqrt{\det\boldsymbol{\Sigma}_{s}}}. \label{CsAA}%
\end{equation}
To compute $\Pi_{s}$ and $\boldsymbol{\Sigma}_{s}$, we note that
$\mathbf{V}_{0}(\mu)$ describes a pure state, which means that its symplectic
spectrum is trivial $\nu_{\pm}^{0}=1$ and $\mathbf{V}_{0}(\mu)=\mathbf{S}%
_{0}\mathbf{S}_{0}^{T}$. Since $G_{s}(1)=\Lambda_{s}(1)=1$, we may write
\begin{equation}
\Pi_{s}:=4G_{1-s}(\nu_{+}^{1})G_{1-s}(\nu_{-}^{1})~, \label{grandePI}%
\end{equation}
and%
\begin{equation}
\boldsymbol{\Sigma}_{s}:=\mathbf{V}(\mu)+\mathbf{S}_{1}\left(
\begin{array}
[c]{cc}%
\Lambda_{1-s}(\nu_{-}^{1})\mathbf{I} & \mathbf{0}\\
\mathbf{0} & \Lambda_{1-s}(\nu_{+}^{1})\mathbf{I}%
\end{array}
\right)  \mathbf{S}_{1}^{T}~. \label{grandeSIGMA}%
\end{equation}

Thus, we only need to compute the symplectic decomposition
\begin{equation}
\mathbf{V}_{1}(\mu,\tau)=\mathbf{S}_{1}\left(
\begin{array}
[c]{cc}%
\nu_{-}^{1}\mathbf{I} & \mathbf{0}\\
\mathbf{0} & \nu_{+}^{1}\mathbf{I}%
\end{array}
\right)  \mathbf{S}_{1}^{T}. \label{CMpartenza}%
\end{equation}
Note that $\mathbf{V}_{1}(\mu,\tau)$ is in the normal form of
Eq.~(\ref{formEASY}), with $a=\tau\mu+1-\tau$, $b=\mu$, and $c=\sqrt{\tau
(\mu^{2}-1)} $. Therefore, from Eq.~(\ref{spectrumEASY}) we derive the
symplectic spectrum $\nu_{-}^{1}=1$ and $\nu_{+}^{1}=\tau+\mu(1-\tau)$. Then,
from Eq.~(\ref{esseEASY}) we derive the following terms%
\begin{equation}
\omega_{-}=\sqrt{\frac{\tau(\mu-1)}{1+\tau+\mu(1-\tau)}},~\omega_{+}%
=\sqrt{\frac{1+\mu}{1+\tau+\mu(1-\tau)}}~,
\end{equation}
for the symplectic matrix $\mathbf{S}_{1}=\mathbf{S}(\tau,\mu)$.

By replacing the previous expressions in Eqs.~(\ref{grandePI})
and~(\ref{grandeSIGMA}), we get%
\begin{equation}
\Pi_{s}:=4G_{1-s}[\tau+\mu(1-\tau)],
\end{equation}
and
\begin{equation}
\boldsymbol{\Sigma}_{s}:=\mathbf{V}(\mu)+\mathbf{S}(\tau,\mu)\left(
\begin{array}
[c]{cc}%
\mathbf{I} & \mathbf{0}\\
\mathbf{0} & \Lambda_{1-s}[\tau+\mu(1-\tau)]\mathbf{I}%
\end{array}
\right)  \mathbf{S}(\tau,\mu)^{T}.
\end{equation}
Now, we can use these expressions of $\Pi_{s}$ and $\boldsymbol{\Sigma}_{s}$
in Eq.~(\ref{CsAA}) and compute the supremum of $P(r,s)$ in Eq.~(\ref{QHBdef}%
). One can easily check that the resulting QHB $H_{\text{quant}}(r)$ satisfies
Eq.~(\ref{QHBquantumEPR2}) for $r\geq2\ln\left[  1+\bar{n}\left(  1-\sqrt
{\tau}\right)  \right]  $, while it can become infinite when $r<2\ln\left[
1+\bar{n}\left(  1-\sqrt{\tau}\right)  \right]  $.

\end{document}